\documentclass[prd,tightenlines,nofootinbib]{revtex4}
\usepackage{epsfig,color,eso-pic}
\usepackage{amsmath,amscd,amssymb}
\usepackage{graphics,bigints}
\usepackage{diagbox}

\newcommand{\imu}{{\rm i}}

\newcommand{\Vek}[1]{{\boldsymbol#1}}

\newcommand{\zr}[1]{\mbox{\hspace*{#1em}}}
\newcommand{\ID}{\mbox{{\sf 1}\zr{-0.16}\rule{0.04em}{1.55ex}\zr{0.1}}}

\usepackage[normalem]{ulem}

\begin{document}

\title{Quantum energies of BPS vortices in $D=2+1$ and $D=3+1$}

\author{N. Graham$^{a)}$, H. Weigel$^{b)}$}

\affiliation{
$^{a)}$Department of Physics, Middlebury College
Middlebury, VT 05753, USA\\
$^{b)}$Institute for Theoretical Physics, Physics Department,
Stellenbosch University, Matieland 7602, South Africa}

\begin{abstract}
We consider vortices in scalar electrodynamics and compute the leading 
quantum correction to their energies for the BPS case of identical
classical masses of the Higgs and gauge fields. In particular, we
focus on the winding number $n$ dependence of these corrections, from
which we can extract the binding energies  of configurations with
larger $n$. For both dimensionalities, $D=2+1$ and $D=3+1$, we find
that quantum corrections are negative and scale approximately linearly
with $n$, so that combined vortices are favored over isolated ones.
\end{abstract}

\maketitle

\section{Introduction}

The Abrikosov-Nielsen-Olesen (ANO) vortex
\cite{Abrikosov:1956sx,ABRIKOSOV1957199,Nielsen:1973cs} provides the simplest 
example of a topological soliton with integer winding number, relevant to 
applications in condensed matter \cite{Tinkham:1996}, particle physics 
\cite{Nambu:1977ag} and cosmology \cite{Kibble:1976sj,Shellard:1994}. 
It arises as a classical solution to the field equations in a theory of scalar
electrodynamics with spontaneous symmetry breaking, where the scalar can be the 
field of Cooper pairs in a superconductor or a Higgs-like field in a particle 
physics model or a domain wall binding a cosmic string. After
spontaneous symmetry breaking, both the scalar and gauge fields have
nonzero mass, and for string configurations that are localized in a
two-dimensional transverse plane, the magnetic flux through the plane
corresponds to a conserved topological winding number. In the
Bogomolny-Prasad-Sommerfeld (BPS) \cite{Bogomolny:1975de,Prasad:1975kr}
case of equal classical masses, which we will focus on here, the classical
energy is directly proportional to this flux.

Given this classical field theory picture, it is natural to ask how these 
results are modified by quantum corrections, and whether the direct
proportionality of energy to winding is maintained. To one loop,
these corrections consist of the vacuum polarization energy (VPE), the
renormalized sum over the zero-point energies $\frac{1}{2} \hbar
\omega$ for small oscillations around the classical
background. Formally, the VPE is defined as
\begin{equation}
\Delta E=
\frac{\hbar}{2}\sum_k \left[\omega_k-\omega_k^{(0)}\right]\Bigg|_{\rm ren.}\,,
\label{eq:General}
\end{equation}
where $\omega_k$ and $\omega_k^{(0)}$ denote the spectra of the quantum
fluctuations with and without the vortex background, respectively. The subscript
indicates that this divergent sum requires renormalization, which is
the primary challenge for the calculation.

Let us briefly explain the renormalization procedure. In the
background of a static localized configuration we express the
renormalized VPE as the sum
\begin{equation}
\Delta E=E_{\rm b.s.}+\frac{1}{2}\int_0^\infty dk\,\sqrt{k^2+M^2}\left(
\Delta\rho(k)-\Delta\rho^{(1)}(k)-\Delta\rho^{(2)}(k)\ldots\right)+
E_{\rm   FD}+E_{\rm CT}\,,
\label{eq:VPE1}\end{equation}
where we have chosen natural units with $\hbar=c=1$.
Here $E_{\rm b.s.}$ is the contribution from the discrete bound states 
in the potential induced by the background. The contribution from the
continuum scattering states is given by the momentum integral,
in which the effect of the background is to change the density of states. We call this change
$\Delta\rho(k)$. It has a Born expansion in the strength of the
potential, and subtracting sufficiently many leading orders of this
expansion renders the momentum integral finite. 
This convergent integral can then be combined 
with $E_{\rm b.s.}$ and analytically continued to the imaginary
momentum axis. The subtracted Born terms are added back in as Feynman diagrams, $E_{\rm FD}$,
which arise from an equivalent expansion of the effective action. When
combined with standard counterterms, $E_{\rm CT}$, the sum 
$E_{\rm FD}+E_{\rm CT}$ is also finite.

We will use scattering theory to compute the change in the density of
states and its Born expansion in a partial wave expansion
\cite{Graham:2022rqk,Graham:2009zz}. Here an
additional complication arises as a result of the gauge field
winding: scattering theory requires that the background fields vanish
at spatial infinity, but string configurations instead approach a 
nontrivial pure gauge, reflecting the topological winding. To remove this
behavior, we use a gauge transformation that makes the fields trivial at
infinity, at the cost of introducing a singularity in the gauge field
at the origin. The associated magnetic field is unchanged, and remains
zero at infinity and finite at the origin. This singularity does not
contribute to the final result since it is a gauge artifact, but
careful regularization is required to implement it consistently 
while maintaining gauge symmetry \cite{Graham:2021mbc}.

The first complete treatment of this problem was given in
Ref.~\cite{Baacke:2008sq}. There, a more {\it ad hoc} scheme was used
to subtract and add back in terms corresponding to both the
renormalization counterterms and the gauge singularity. While in
principle the calculations should be equivalent, our approach provides 
a more systematic separation of the divergences, in the process demonstrating 
that, surprisingly, a much larger number of partial waves are needed to 
obtain the large $k$ behavior of $\Delta\rho(k)$ that is consistent with 
that obtained from analyzing Feynman diagrams. This effect explains some of 
the discrepancies between our results and previous calculations, some of which 
appeared to converge without renormalization
\cite{Alonso-Izquierdo:2016bqf,Pasipoularides:2000gg}
when too few partial waves are taken into account; including a
larger number of partial waves restores the expected divergence.
Other discrepancies arise from the peculiarities of the renormalization 
conditions, which we choose to be on-shell.

In this paper we consider vortices in both $D=2+1$ and $D=3+1$ spacetime
dimensions. In the former case, the lower dimension means fewer
diagrams are divergent, as is typical in quantum field theory.
We nonetheless include finite counterterms to implement the same
on-shell renormalization conditions as in $3+1$ dimensions, in both
cases ensuring that the residues of the propagator poles for both
particles, corresponding to the normalization of single-particle
states, are left unchanged, as is the pole location for the Higgs
particle, corresponding to its mass.  The mass of the gauge particle is
corrected by  quantum effects, so in the end the theory is specified
by the two masses, or equivalently by the Higgs mass and the gauge
coupling constant.  However, while this mass splitting occurs at
one-loop order, its effects on the VPE enter at two loops and can be
ignored in our calculation.  In the case of $3+1$ dimensions, the
scattering density of states remains the same as in $D=2+1$, but we
must use analytic continuation to consistently include the integral
over the momentum in the trivial direction \cite{Graham:2001dy}.

Since the classical BPS vortex has energy proportional to its winding
number, the classical energy of a winding $n$ vortex is equal to the
energy of $n$ isolated vortices. In the condensed matter system, it
represents the boundary between Type I and Type II superconductors.
The quantum correction will therefore either stabilize or
destabilize the higher winding configurations; by carrying out the
calculation through $n=4$ we find that higher winding is stabilized.

Throughout the paper we treat the $D=2+1$ and $D=3+1$ cases in
parallel without introducing separate notations for the most part, and
use the context to identify the particular case.  In Section II we
introduce the classical vortex configuration in singular gauge. The
quantization of the theory at one loop and the corresponding on-shell
renormalization procedure are described in Sections III and IV,
respectively. In Section V we explain the computation of the VPE using
scattering data on the imaginary momentum axis and show how we move
the divergent contributions from the momentum integral into Feynman
diagrams, which are then combined with the counterterms from Section
IV. Numerical results are presented and discussed in Section VI, and we
give a short summary and conclusion in Section VII. In a short Appendix we 
estimate the higher-order effects of different masses for the VPE.

\section{Classical solutions}

We start from the Lagrangian
\begin{equation}
\mathcal{L}=-\frac{1}{4}F_{\mu\nu}F^{\mu\nu}+|D_\mu\Phi|^2
-\frac{\lambda}{4}\left(|\Phi|^2-v^2\right)^2\,,
\label{eq:lag1}\end{equation}
where as usual $F_{\mu\nu}=\partial_\mu A_\mu-\partial_\nu A_\mu$
and $D_\mu\Phi=\left(\partial_\mu-\imu e A_\mu\right)\Phi$ for an
Abelian gauge theory.

In singular gauge, the profiles associated with winding number $n$ are
the functions $g(\rho)$ and $h(\rho)$ within the ans\"atze
\begin{equation}
\Phi_S=vh(\rho) 
\qquad {\rm and}\qquad 
\Vek{A}_S=nv\hat{\Vek{\varphi}}\,\frac{g(\rho)}{\rho}\,,
\label{eq:string1}\end{equation}
where $\rho=evr$ is dimensionless while $r$ is the physical radial
coordinate. Here $g(\rho)$ ranges from $0$ to $1$ and $h(\rho)$ ranges
from $1$ to $0$ as $\rho$ goes from $0$ to $\infty$.

This field configuration leads to the energy functional
\begin{equation}
E_{\rm cl}=2\pi v^2\int_0^\infty \rho d\rho\,
\left[\frac{n^2}{2}\frac{g^{\prime2}}{\rho^2}
+h^{\prime2}+n^2\frac{h^2}{\rho^2}g^2
+\frac{\lambda}{4e^2}\left(h^2-1\right)^2\right]\,,
\label{eq:ecl}\end{equation}
where primes denote derivatives with respect to $\rho$. In $D=2+1$,
where $v^2$ has dimensions of mass,  $E_{\rm cl}$ is the vortex
energy, while in $D=3+1$, where $v$ has dimensions of mass, it is the
energy per unit length of the vortex.

Recalling the tree level mass relations $M_H^2=\lambda v^2$ and $M_A^2=2v^2e^2$, the 
coefficient of the last terms becomes
$$
\frac{\lambda}{4e^2}=\frac{M_H^2}{2M_A^2}\,,
$$
so that when measured in units of $2\pi v^2$, the classical energy
only depends on the ratio of the two masses.

\begin{figure}
\centerline{\epsfig{file=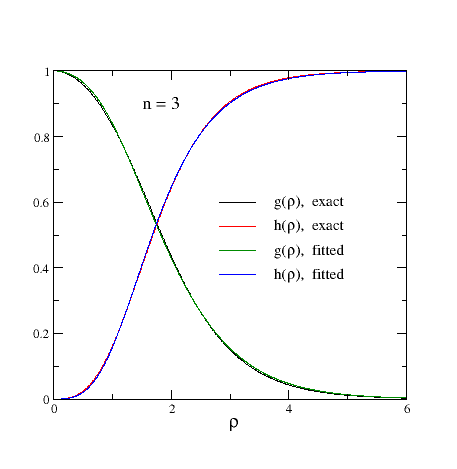,width=7cm,height=3cm}~~~
\epsfig{file=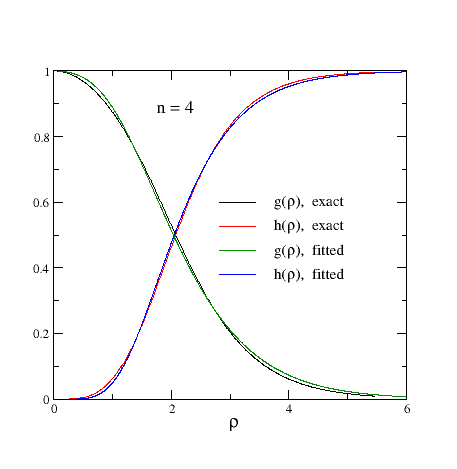,width=7cm,height=3cm}}
\caption{\label{fig:class}Comparison of exact and fitted profile functions (from
Eq.~(\ref{eq:profilefit}) and Table \ref{tab:fit}) for winding numbers $n=3$ and $n=4$.}
\end{figure}

In the BPS case, which we assume henceforth, the coupling constants
are related by $\lambda=2e^2$, {\it i.e} $M_H=M_A=M$. Then the energy
functional can be written as sums of non-negative quantities plus a
surface contribution
\begin{equation}
E_{\rm cl}=2\pi v^2\int_0^\infty \rho d\rho\,\left\{
\frac{1}{2}\left[\frac{n}{\rho}g^\prime-(h^2-1)\right]^2
+\left[h^\prime-\frac{n}{\rho}gh\right]^2\right\}
+2\pi nv^2g(h^2-1)\Big|_0^\infty\,.
\label{eq:EBPS}\end{equation}
Hence the minimal energy is fully determined by the winding number,
yielding $E_{\rm cl}=2\pi nv^2=2\pi n\frac{M^2}{\lambda}$, with the 
corresponding profiles obeying the first-order differential equations
\begin{equation}
g^\prime=\frac{\rho}{n}(h^2-1)
\qquad {\rm and}\qquad
h^\prime=\frac{n}{\rho}gh\,,
\label{eq:cldeqBPS}\end{equation}
with the boundary conditions
\begin{equation}
h(0)= 1 - g(0)=0 \qquad{\rm and}\qquad
\lim_{\rho\to\infty}h(\rho) =
1-\lim_{\rho\to\infty}g(\rho)=1\,,
\label{eq:bcANO}\end{equation}
which correspond to field configurations that approach constant vacuum values
at spatial infinity. The topological structure appears through a
singularity in the gauge field at the origin, while the magnetic field
remains finite everywhere. 
We have solved the differential equations in~(\ref{eq:cldeqBPS})
numerically, but for later use in the scattering calculation an
approximate expression in terms of elementary functions is very
helpful. It turns out that for $1\le n\le4$ the correlation
coefficients for the fit
\begin{equation}
h(\rho)=\alpha_2\tanh^n(\alpha_1\rho)+[1-\alpha_2]\tanh^n(\alpha_0\rho)
\qquad{\rm and}\qquad
g(\rho)=\beta_1\rho\,\frac{1-\tanh^2(\beta_2\rho)}{\tanh(\beta_1\rho)}
\label{eq:profilefit}
\end{equation}
with the fit parameters $\alpha_i$ and $\beta_i$ listed in Table~\ref{tab:fit} deviate from
unity by~$10^{-4}$ or less from the numerical solutions to Eq.~(\ref{eq:cldeqBPS}). The quality
of the parameterization is also reflected by the smallness of the relative error
$\delta E=E_{\rm fit}/(2\pi n v^2)-1$, which is also presented in Tab.~\ref{tab:fit}.
\begin{table}[ht]
{\begin{tabular}{c|ccc|cc|c}
$n$ & $\alpha_0$ & $\alpha_1$ & $\alpha_2$ & $\beta_1$ & $\beta_2$ & $\delta E$ \cr
\hline
1 & 0.8980 & 0.6621 & 0.1890 & 0.5361 & 0.7689 & $4.2\times10^{-6}$ \\
2 & 0.9072 & 0.8288 & 2.6479 & 1.0949 & 0.8042 & $1.9\times10^{-5}$ \\
3 & 0.8290 & 0.7882 & 5.1953 & 1.1328 & 0.7425 & $1.5\times10^{-3}$ \\
4 & 0.7755 & 0.7350 & 5.2009 & 1.1034 & 0.6853 & $5.6\times10^{-3}$
\end{tabular}}
\caption{\label{tab:fit}Fit parameters for vortex profiles. 
The quality of the fit is estimated by the accuracy of the 
energy: $\delta E=E_{\rm fit}/(2\pi n v^2)-1$, where $E_{\rm fit}$ is obtained
by substituting (\ref{eq:profilefit}) into Eq.~(\ref{eq:ecl}).}
\end{table}
This accuracy test indicates that the agreement is excellent for $n=1$ and $n=2$, 
but merely good for $n=3$ and $n=4$. To show that the fit is nevertheless 
also suitable in these cases, we display the corresponding profiles
in Figure \ref{fig:class}. A graphical comparison for $n=1$ and $n=2$
does not provide visible differences and we refrain from its presentation.

\section{Quantization}

In this Section we quantize the theory, including the relevant ghost fields 
required for gauge fixing, and derive the equations of motion for the harmonic 
fluctuations. We will construct the renormalization counterterms in the next
Section.

\subsection{Lagrangian for fluctuations}

We introduce time-dependent fluctuations about the vortex via
\begin{equation}
\Phi=\Phi_S+\eta \qquad {\rm and}\qquad 
A^\mu=A_S^\mu+a^\mu
\label{eq:fluct}\end{equation}
and extract the harmonic terms after several integrations by parts as
\begin{align}
\mathcal{L}^{(2)}&=-\frac{1}{2}\left(\partial_\mu a_\nu\right)\left(\partial^\mu a^\nu\right)
+\frac{1}{2}\left(\partial_\mu a^\mu\right)^2+|\Phi_S|^2a_\mu a^\mu\cr
&\hspace{0.5cm}
+|D_\mu\eta|^2-\left(|\Phi_S|^2-1\right)|\eta|^2
-\frac{1}{2}\left(\Phi_S\eta^\ast+\Phi^\ast_S\eta\right)^2\cr
&\hspace{0.5cm}
+\imu\left(\Phi_S\eta^\ast-\Phi^\ast_S\eta\right)\partial_\mu a^\mu
+2\imu a^\mu\left(\eta^\ast D_\mu\Phi_S-\eta D^\ast_\mu\Phi^\ast_S\right)a^\mu\,,
\label{eq:lag2}\end{align}
where $D_\mu$ is the covariant derivative with the vortex
configuration $\Vek{A}_S$ substituted, $D_0=\partial_t$
and $\Vek{D}=\Vek{\nabla}+\imu n\Vek{\varphi}\frac{g(\rho)}{\rho}$, and we
have chosen units with $ev=1$ such that both particles have classical
mass $\sqrt{2}$. The gauge is fixed from the vortex background by
adding an $R_\xi$ type Lagrangian with $\xi=1$ to cancel the
$\eta \partial_\mu a^\mu$ term,
\begin{equation}
\mathcal{L}_{\rm gf}=-\frac{1}{2}\left[\partial_\mu a^\mu
+\imu\left(\Phi_S\eta^\ast-\Phi_S^\ast\eta\right)\right]^2\,.
\label{eq:laggf}\end{equation}
Collecting the harmonic terms yields
\begin{align}
\mathcal{L}^{(2)}+\mathcal{L}_{\rm gf}
&=-\frac{1}{2}\left(\partial_\mu a_\nu\right)\left(\partial^\mu a^\nu\right)
+|\Phi_S|^2a_\mu a^\mu
+|D_\mu\eta|^2-\left(3|\Phi_S|^2-1\right)|\eta|^2
+2\imu a^\mu\left(\eta^\ast D_\mu\Phi_S-\eta D^\ast_\mu\Phi^\ast_S\right)\,.
\label{eq:lag3}\end{align}

We still have to subtract the ghost contribution to the VPE associated
with the gauge fixing in Eq.~(\ref{eq:laggf}), which we write as
$\mathcal{L}_{\rm gf}=-\frac{1}{2}G^2$.  The infinitesimal gauge
transformations reads
\begin{equation}
A^\mu\to A^\mu+\partial \chi\,, \qquad 
\Phi_0+\eta\to \Phi_0+\eta+\imu\chi(\Phi_0+\eta)
\qquad \mbox{so that} \qquad \eta\to\eta +\imu\chi(\Phi_0+\eta)\,.
\label{eq:gt1}\end{equation}
Then 
\begin{equation}
\frac{\partial G}{\partial \chi}\Big|_{\chi=0}
=\partial_\mu\partial^\mu+\left(2|\Phi_0|^2+\Phi_0\eta^\ast+\Phi_0^\ast\eta\right)\,.
\label{eq:fp1}\end{equation}
This induces the ghost Lagrangian (in agreement with Refs. \cite{Rebhan:2004vu,Lee:1994pm})
\begin{equation}
\mathcal{L}_{\rm gh}=-\partial_\mu\overline{c}\,\partial^\mu c + 2|\Phi_0|\overline{c}c 
+\mbox{non-harmonic terms}\,.
\label{eq:fp2}\end{equation}
Its VPE is (the negative of) that of a Klein Gordon field of mass $\sqrt{2}$ 
in the background potential $2(h^2-1)$, which can be easily
computed. Since it is a complex ghost field, it must be subtracted
with a factor of two from the above. Note that from  
Eq.~(\ref{eq:lag3}), the non-transverse components of $a_\mu$ couple to
the same background potential. 

In $D=2+1$ the spectrum consists of four real decoupled fields with mass
$\sqrt{2}$: $a_1$, $a_2$, ${\sf Re}(\eta)$ and ${\sf Im}(\eta)$. Three other
fields, also with mass $\sqrt{2}$, are fully decoupled: $a_0$ and the two ghosts. 
The ghosts count negatively, and one of them cancels against the temporal component of 
the gauge field since they obey the same equation of motion. In total
there are then $5-2=3$ physical degrees of freedom. When computing the
VPE, we thus have to subtract a boson type contribution from the background $2(h^2-1)$. 
In $D=3+1$ the gauge field has an additional decoupled longitudinal
component, so the non-transverse and ghost contributions cancel
completely and we only need to  consider $a_1$ and $a_2$ together with
the complex Higgs field.

\subsection{Wave-equations for quantum fluctuations}

To formulate the scattering problem, we employ a partial wave decomposition using the
complex combinations 
\begin{align}
a_x+\imu a_y=\sqrt{2}\imu{\rm e}^{-\imu\omega t}\sum_\ell a_\ell(\rho){\rm e}^{\imu \ell \varphi}
\quad {\rm and}\quad
\eta={\rm e}^{-\imu\omega t}\sum_\ell\eta_\ell(\rho){\rm e}^{\imu \ell \varphi}\,.
\label{eq:partialwave}
\end{align}
We have an analogous expansion for $a_x-\imu a_y$ and $\eta^\ast$ and, in general, a coupled 
system of four radial functions. In the BPS case, fortunately, this system decouples into two 
sub-blocks, with the one containing $a_x-\imu a_y$ and $\eta^\ast$ being identical to the above. 
It will thus suffice to solve
\begin{align}
\frac{1}{\rho}\frac{\partial}{\partial\rho}\rho
\frac{\partial}{\partial\rho} \eta_\ell(\rho)&=-q^2\eta_\ell(\rho)
+\left[\frac{\ell^2-2n\ell g(\rho)+n^2g^2(\rho)}{\rho^2}+3(h^2(\rho)-1)\right]\eta_\ell(\rho)
+\sqrt{2}d(\rho)a_{\ell+1}(\rho)\cr
\frac{1}{\rho}\frac{\partial}{\partial\rho}\rho
\frac{\partial}{\partial\rho}a_{\ell+1}(\rho)&=-q^2a_{\ell+1}(\rho)
+\left[\frac{(\ell+1)^2}{\rho^2}+2(h^2(\rho)-1)\right]a_{\ell+1}(\rho)
+\sqrt{2}d(\rho)\eta_\ell(\rho)
\label{eq:fluct1}\end{align}
with $q^2=\omega^2-2$ and double its VPE. Here the off-diagonal coupling is 
$d(\rho)=\frac{\partial h(\rho)}{\partial\rho}+\frac{n}{\rho}h(\rho)g(\rho)$.

Finally, the ghost field fully decouples and has a partial wave
expansion analogous to the Higgs field in Eq.~(\ref{eq:partialwave}),
\begin{equation}
\frac{1}{\rho}\frac{\partial}{\partial\rho}\rho
\frac{\partial}{\partial\rho} \zeta_\ell(\rho)
=-q^2\zeta_\ell(\rho)+\left[\frac{\ell^2}{\rho^2}+2(h^2(\rho)-1)\right]\zeta_\ell(\rho)\,.
\label{eq:fluc2}\end{equation}

\section{Renormalization}

In this Section we describe the renormalization of the one-loop corrections arising
from the fluctuations about the vortex. We begin by analyzing the effective
action.

\subsection{Effective actions in $D=2+1$ and $D=3+1$}

To identify the ultraviolet divergences in the form of Feynman diagrams
\cite{Becchi:1974md,Irges:2017ztc}, we consider the Lagrangian
\begin{equation}
\mathcal{L}=\frac{1}{2}\left(\partial_\mu \phi\right)\left(\partial^\mu \phi^{\rm t}\right)
-\frac{1}{2} \phi M^2 \phi^{\rm t} - \phi V \phi^{\rm t}
\label{eq:LagFD}
\end{equation}
with four real fields $\phi=(\eta_1,\eta_2,a_x,a_y)$. The Cartesian components of the 
gauge fields have been defined in Eq.~(\ref{eq:partialwave}) above, while 
$\eta=\left(\eta_1+\imu\eta_2\right)/\sqrt{2}$. We then  Taylor expand the effective 
action for these real scalar fields as
\begin{align}
\mathcal{A}_{\rm eff}&=\frac{\imu}{2}{\rm Tr}\,{\rm Log}\left[\partial^2+M^2-\imu0^{+}+2V\right]\cr
&=\mathcal{A}^{(0)}_{\rm eff}
+\imu{\rm Tr}\left[\hat{G}V\right]
-\imu{\rm Tr}\left[\hat{G}V\hat{G}V\right]
+\frac{4\imu}{3}\,{\rm Tr}\left[\hat{G}V\hat{G}V\hat{G}V\right]
-2\imu{\rm Tr}\left[\hat{G}V\hat{G}V\hat{G}V\hat{G}V\right]+\ldots,
\label{eq:aeff}\end{align}
where $\hat{G}=\left(\partial^2+M^2-\imu0^+\right)^{-1}$ times the $4\times4$ unit matrix.
The functional trace is over the space-time coordinates as well as the
elements of $\phi$ and the ellipsis in Eq.~(\ref{eq:aeff}) represents
ultra-violet finite terms. The potential matrix is given by
$V=V_0+V_1+V_2$, with
\begin{equation}
V_0=e^2\begin{pmatrix}\frac{3}{2}\left(\Phi^2_S-v^2\right) &
0 & \sqrt{2}\,\hat{\Vek{x}}\cdot\Vek{A}_S\Phi_S & 
\sqrt{2}\,\hat{\Vek{y}}\cdot\Vek{A}_S\Phi_S\\[1mm]
0 & \frac{3}{2}\left(\Phi^2_S-v^2\right) &
-(\sqrt{2}/e)\,\hat{\Vek{x}}\cdot\Vek{\nabla} \Phi_S &
-(\sqrt{2}/e)\,\hat{\Vek{y}}\cdot\Vek{\nabla} \Phi_S\\[1mm]
\sqrt{2}\,\hat{\Vek{x}}\cdot\Vek{A}_S\Phi_S & 
-(\sqrt{2}/e)\,\hat{\Vek{x}}\cdot\Vek{\nabla} \Phi_S & 
\left(\Phi^2_S-v^2\right) & 0  \\[1mm]
\sqrt{2}\,\hat{\Vek{y}}\cdot\Vek{A}_S\Phi_S  & 
-(\sqrt{2}/e)\,\hat{\Vek{y}}\cdot\Vek{\nabla}\Phi_S &
0 & \left(\Phi^2_S-v^2\right)
\end{pmatrix}
\label{eq:PotMatrix0}
\end{equation}
and
\begin{equation}
V_1=e\begin{pmatrix} 0 & 1 & 0 & 0 \\[0mm] -1 & 0 & 0 & 0 \\[0mm]
0 & 0 & 0 & 0 \\[0mm] 0 & 0 & 0 & 0\end{pmatrix}\, \Vek{A}_S\cdot\Vek{\nabla}
\qquad {\rm and}\qquad
V_2=\frac{e^2}{2}\begin{pmatrix} 1 & 0 & 0 & 0 \\[0mm] 0 & 1 & 0 & 0 \\[0mm]
0 & 0 & 0 & 0 \\[0mm] 0 & 0 & 0 & 0\end{pmatrix}\, \Vek{A}_S\cdot\Vek{A}_S
\label{eq:PotMatrix1}
\end{equation}
Here we have separated out $V_1$ and $V_2$ because they relate to the singular terms in the
scattering problem, while $V_0$ is the $4\times4$ representation of the non-singular terms
in Eq.~(\ref{eq:fluct1}).  The renormalization program via Feynman diagrams in dimensional 
regularization is carried out with the full potential matrix $V$, while the subtractions 
that we have indicated in Eq.~(\ref{eq:VPE1}) should only involve $V_0$ supplemented by the 
wave-function renormalization of the gauge boson, which is simplified by the fake boson trick 
described below.

\subsection{On-shell renormalization counterterms}

The counterterm Lagrangian has four terms,
\begin{equation}
\mathcal{L}_{\rm CT}= C_g F_{\mu\nu} F^{\mu\nu} + C_h \left|D_\mu\Phi\right|^2
+C_0\left(\Phi^2-v^2\right)+C_V\left(\Phi^2-v^2\right)^2\,.
\label{eq:Lct}\end{equation}
The $C_0$ counterterm arises from varying the vacuum expectation value $v$ in the 
original Lagrangian, Eq.~(\ref{eq:lag1}). The coefficient is chosen such that
it exactly cancels $\imu{\rm Tr}\left[\hat{G}V_0\right]$ in
Eq.~(\ref{eq:aeff}), which is 
the no-tadpole condition. The coefficients $C_g$ and $C_h$ are determined such that the 
residues of the propagators at the respective masses have
no quantum correction, while $C_V$ is fixed such that the pole location of the 
Higgs propagator (which determines its mass) does not change at one-loop order.
The pole location of the gauge field propagator is then an output of the calculation, 
which can be expressed in terms of the other Lagrangian parameters.

In what follows, $D$ will be the physical dimension while $D_\epsilon$ 
is its continuation in
dimensional regularization. That is, for $D=3+1$ we have $D_\epsilon=4-2\epsilon$
and $\epsilon\,\mbox{\small$\searrow$}\,0$.
We use lower-case letters to denote finite counterterm coefficients; where counterterms 
diverge, we use the corresponding upper-case letter to denote the coefficients including
a divergent part. 

To determine the mass and wave-function renormalization we need to expand the effective action 
to quadratic order in the Higgs and gauge fields. Let us first discuss $\mathcal{A}_{\rm A}$, the 
contributions to the effective action that are quadratic in the gauge
field and superficially quadratically 
divergent, before imposing gauge invariance of the regulator.  These
contributions arise from the terms linear in $V_2$ and quadratic in $V_1$,
\begin{equation}
\mathcal{A}_{\rm A}=C_G\int d^{D}x\, F_{\mu\nu}F^{\mu\nu}+
\imu {\rm Tr}\left[\hat{G}V_2\right]
-\imu {\rm Tr}\left[\hat{G}V_1\hat{G}V_1\right] \,,
\label{eq:AeffA}
\end{equation}
which yields
\begin{align}
\mathcal{A}_{\rm A}&=C_G\int d^{D}x\, F_{\mu\nu}F^{\mu\nu}
\cr & \hspace{1cm}
+e^2\frac{\mu^{3-D_\epsilon}}{\left(4\pi\right)^{D_\epsilon/2}}\Gamma\left(1-\frac{D_\epsilon}{2}\right)
\int \frac{d^Dk}{(2\pi)^D}\,\widetilde{A}_\mu(k)\widetilde{A}^\mu(-k)
\int_0^1 dx\left[M^{D_\epsilon-2}-\left(M^2-x(1-x)k^2\right)^{D_\epsilon/2-1}\right]\,,
\label{eq:AeffA1}
\end{align}
where $\widetilde{A}^\mu(k)$ denotes the Fourier transform of the gauge field.
We have repeatedly used $p_\mu\widetilde{A}^\mu(k)=0$, which results
from the vortex property $\partial_\mu A^\mu(x)=0$ and also implies
\begin{equation}
\int d^Dx F_{\mu\nu}F^{\mu\nu}=
2\int \frac{d^Dk}{(2\pi)^D}\, k^2\widetilde{A}_\mu(k)\widetilde{A}^\mu(-k)\,.
\label{eq:Ftensor1}\end{equation}
We may also assume $p_\mu\widetilde{A}^\mu(k)=0$ generally when determining the 
wave-function renormalization $C_g$ because it still allows us to uniquely identify
the field strength tensor in the quadratic expansion of the effective action.\footnote{If 
there was parity violation, the dual field strength tensor would 
also contribute and the assumption would not be justified.}
The residue of the $D=2$ pole is zero, so the quadratic divergence 
disappears and we can continue to the dimension of interest.
There are additional superficial divergences in $D=3+1$ when expanding the
effective action, Eq.~(\ref{eq:aeff}). However, the logarithmic divergences
from $V_0\otimes V_2$ and $V_0\otimes V_1\otimes V_1$ cancel, as do those
from $V_2\otimes V_2$, $V_2\otimes V_1\otimes V_1$ and $V_1\otimes
V_1\otimes V_1\otimes V_1$. Hence the quadratic order in $V$ is sufficient to 
implement renormalization.

To collect all terms that are quadratic in the fields we introduce $\widetilde{v}_H(k)$ 
and $\widetilde{a}(k)$ as the Fourier transforms of
\begin{equation}
v_H=\Phi_S^2-v^2
\qquad {\rm and} \qquad
a=\sqrt{2}e\begin{pmatrix}\hat{e\Vek{x}}\cdot\Vek{A}_S\Phi_S
& e\hat{\Vek{y}}\cdot\Vek{A}_S\Phi_S \cr
-\hat{\Vek{x}}\cdot\Vek{\nabla}\Phi_S & 
-\hat{\Vek{y}}\cdot\Vek{\nabla}\Phi_S\end{pmatrix}\,,
\label{eq:FT1}\end{equation}
respectively. For example,
$\widetilde{v}_H(k)=\int d^4x\,\left[\Phi_S^2-v^2\right]{\rm e}^{\imu k_\mu x^\mu}$.
In $D=2+1$ we then have
\begin{align}
\mathcal{A}^{(3)}&=\int d^3x\,\left[c_gF_{\mu\nu}F^{\mu\nu}
+c_h\left(D_\mu \Phi\right)^\ast\left(D^\mu \Phi\right)
+c_v\left(\Phi\Phi^\ast-v^2\right)^2\right]\cr
&\hspace{0.5cm}+\frac{1}{8\pi M}\int \frac{d^3k}{(2\pi)^3}
\left[\frac{11e^4}{2}\widetilde{v}_H(k)\widetilde{v}_H(-k)
+2{\rm tr}\left(\widetilde{a}(k)\widetilde{a}^{\rm t}(-k)\right)\right]
\int_0^1\frac{dx}{\sqrt{1-x(1-x)k^2/M^2}}\cr
&\hspace{0.5cm}-\frac{e^2M}{4\pi}\int \frac{d^3k}{(2\pi)^3}
\widetilde{F}_{\mu\nu}(k)\widetilde{F}^{\mu\nu}(-k)\frac{1}{k^2}
\int_0^1dx\,\left[1-\sqrt{1-x(1-x)k^2/M^2}\right]+\ldots\,,
\label{eq:A2D3}
\end{align}
where the ellipsis refer to terms of cubic and higher order.
In $D=3+1$ these contributions have divergences, so we write
\begin{equation}
c_g=C_g-\frac{e^2}{192\pi^2}\left(C_\epsilon+1\right)\,,\quad
c_h=C_h-\frac{e^2}{4\pi^2}C_\epsilon\,,\quad
c_v=C_V+\frac{13e^4}{32\pi^2}C_\epsilon\,,
\end{equation}
with
$C_\epsilon=\frac{1}{\epsilon}-\gamma+
\ln\left(4\pi\frac{\Lambda^2}{M^2}\right)$
from dimensional regularization with scale $\Lambda$. The second-order
effective action is then 
\begin{align}
\mathcal{A}^{(4)}&=\int d^4x\,\left[c_gF_{\mu\nu}F^{\mu\nu}
+c_h\left(D_\mu \Phi\right)^\ast\left(D^\mu \Phi\right)
+c_v\left(\Phi\Phi^\ast-v^2\right)^2\right]\cr
&\hspace{0.5cm}-\frac{1}{8\pi^2}\int \frac{d^4k}{(2\pi)^4}
\left[\frac{13e^4}{4}\widetilde{v}_H(k)\widetilde{v}_H(-k)
+{\rm tr}\left(\widetilde{a}(k)\widetilde{a}^{\rm t}(-k)\right)\right]
\int_0^1dx\,{\rm ln}\left[1-x(1-x)k^2/M^2\right]\cr
&\hspace{0.5cm}-\frac{1}{2}\left(\frac{eM}{4\pi}\right)^2
\int \frac{d^4k}{(2\pi)^4}
\widetilde{F}_{\mu\nu}(k)\widetilde{F}^{\mu\nu}(-k)\frac{1}{k^2}
\int_0^1dx\,\left[1-x(1-x)k^2/M^2\right]
{\rm ln}\left[1-x(1-x)k^2/M^2\right]+\ldots\,.\quad
\label{eq:A2D4}
\end{align}
Here all counterterm coefficients are finite and thus written in lower case.
Comparing the $D=2+1$ and $D=3+1$ cases, a change in the relative coefficients 
between the $\widetilde{v}_H(k)\widetilde{v}_H(-k)$ and 
$\widetilde{a}(k)\widetilde{a}^{\rm t}(-k)$ is observed. This change results from
the ghost contribution in $D=2+1$.

Next, we write the part of the action which is quadratic in the fields
(including the tree level counterterms) as 
\begin{equation}
\mathcal{A}^{(D)}=\int \frac{d^Dk}{(2\pi)^D}\left[
G^{(D)}_H\widetilde{h}(k)\widetilde{h}(-k)
+G^{(D)}_A\widetilde{A}_\mu(k)\widetilde{A}^\mu(-k)\right]\,,
\end{equation}
where $\widetilde{h}(k)$ is the Fourier transform of $h(x)=\Phi(x)-v$. In general (but not for the vortex) 
there are also terms like $\widetilde{h}(k)k_\mu
\widetilde{A}^\mu(-k)$, but they need not to be considered  for our
renormalization conditions, because they would determine the
renormalized coupling. For the Higgs part we find, using $M=\sqrt{2}ev$,
\begin{align}
G_H^{(3)}(k^2)&=\left(\frac{1}{2}+c_h\right)k^2-\left(\frac{M^2}{2}-4v^2c_v\right)
+\frac{e^2}{8\pi M}\left(11M^2-4k^2\right)
\int_0^1\frac{dx}{\sqrt{1-x(1-x)k^2/M^2}}\cr
G_H^{(4)}(k^2)&=\left(\frac{1}{2}+c_h\right)k^2-\left(\frac{M^2}{2}-4v^2c_v\right)
-\frac{e^2}{16\pi^2}\left(13M^2-4k^2\right)
\int_0^1dx\,{\rm ln}\left[1-x(1-x)k^2/M^2\right]\,.
\label{Higgs2}\end{align}
The gauge part is obtained from setting $\Phi=v$,
\begin{align}
G_A^{(3)}(k^2)&=-\frac{1}{2}\left(1-4c_g\right)k^2+\left(1+c_h\right)\frac{M^2}{2}
-\frac{e^2M}{4\pi}\int_0^1 \frac{dx}{\sqrt{1-x(1-x)k^2/M^2}}\cr
&\hspace{3cm}
-\frac{e^2M}{2\pi}\int_0^1 dx\left[1-\sqrt{1-x(1-x)k^2/M^2}\right]\cr
G_A^{(4)}(k^2)&=-\frac{1}{2}\left(1-4c_g\right)k^2+\left(1+c_h\right)\frac{M^2}{2}
+2\left(\frac{eM}{4\pi}\right)^2\int_0^1dx\,{\rm ln}\left[1-x(1-x)k^2/M^2\right]\cr
&\hspace{3cm}
-\left(\frac{eM}{4\pi}\right)^2\int_0^1dx\,\left[1-x(1-x)k^2/M^2\right]
{\rm ln}\left[1-x(1-x)k^2/M^2\right]
\label{gauge2}\end{align}

The on-shell renormalization conditions for the Higgs field are
\begin{equation}
G_H^{(D)}(M^2)=0\qquad {\rm and}\qquad
\frac{\partial G_H^{(D)}(k^2)}{\partial k^2}\Big|_{k^2=M^2}=\frac{1}{2}\,.
\end{equation}
from which we obtain for $D=2+1$
\begin{equation}
c_h=\frac{e^2}{48\pi M}\left[45\ln(3)-28\right]
\qquad {\rm and}\qquad
c_v=\frac{e^4}{96\pi M}\left[28-87\ln(3)\right]\,,
\label{CHD3}\end{equation}
and for $D=3+1$ we find
\begin{equation}
c_h=\frac{e^2}{16\pi^2}\left[17-\frac{10\pi}{\sqrt{3}}\right]
\qquad {\rm and}\qquad
c_v=\frac{e^4}{32\pi^2}\left[19\frac{\pi}{\sqrt{3}}-35\right]\,.
\label{CHD4}\end{equation}

The gauge field renormalization conditions are
\begin{equation}
G_A^{(D)}(M_A^2)=0\qquad {\rm and}\qquad
\frac{\partial G_A^{(D)}(k^2)}{\partial k^2}\Big|_{k^2=M_A^2}=-\frac{1}{2}\,.
\end{equation}
From this we determine the unknowns  $c_g$ and $M_A$, where the latter is
written in terms of the ratio $\mu=\frac{M_A}{M}$. We obtain for $D=2+1$ 
\begin{equation}
c_g(\mu)=\frac{e^2}{32\pi M}\left[\frac{2}{4-\mu^2}
+\frac{1}{\mu}{\rm atanh}\left(\frac{\mu}{2}\right)\right]
\label{cgD3}\end{equation}
and for $D=3+1$
\begin{equation}
c_g(\mu)=-\frac{e^2}{288\pi^2}\frac{1}{\mu^3}\left[\mu\left(\mu^2-12\right)\mu
+3\frac{\mu^4-2\mu^2+16}{\sqrt{4-\mu^2}}
{\rm asin}\left(\frac{\mu}{2}\right)\right] \,.
\label{cgD4}\end{equation}
Within the one-loop order $c_i=\mathcal{O}(\hbar)$ and
$\mu^2-1=\mathcal{O}(\hbar)$ so that $c_g(\mu)-c_g(1)=\mathcal{O}(\hbar^2)$.

Finally, we combine these results to compute the mass ratio $\mu$ in
terms of the Lagrangian parameter $e$.  We obtain for $D=2+1$
\begin{equation}
\frac{2\pi M}{e^2}\left(\mu^2-1\right)=\frac{15}{8}\ln(3)-\frac{7}{6}
+\frac{\mu^2}{2\left(4-\mu^2\right)}-1-\frac{\mu}{4}{\rm atanh\left(\frac{\mu}{2}\right)}
\label{eq:consit2}\end{equation}
and $D=3+1$
\begin{equation}
\frac{8\pi^2}{e^2}\left(\mu^2-1\right)=\frac{9}{2}-\frac{5\pi}{\sqrt{3}}-\frac{\mu^2}{6}
+\frac{16-2\mu^2}{\mu\sqrt{4-\mu^2}}{\rm
  asin}\left(\frac{\mu}{2}\right) \,.
\label{eq:consit3}\end{equation}
At one-loop order to the VPE it is fully consistent to use $\mu=1$ in the finite parts of the counterterm
coefficients. Nevertheless, in Appendix \ref{appA} we will briefly discuss effects resulting from
$c_g(\mu)-c_g(1)\ne0$.

\subsection{Fake boson subtraction}

As seen from Eq.~(\ref{eq:VPE1}), the computation of the VPE requires
us to
move divergent terms from the non-perturbative contribution (which is obtained from
scattering data) to Feynman diagrams. For the remaining logarithmic divergence in 
Eq.~(\ref{eq:AeffA1}) with $D=3+1$, this is not possible for the vortex configuration
since neither its Fourier transform nor its Born approximation to the scattering data
exist. We instead introduce the fake boson technique, in 
which we begin by considering bosonic fluctuations about the static potential $V(x)=V_f(r)$. 
We take this boson field to be complex because its scattering data will later be combined 
with those from the complex vortex fluctuations in Eq.~(\ref{eq:fluct1}). The second-order 
effective action for this boson field is
\begin{align}
\mathcal{A}_{\rm (fb)}&=
\frac{1}{2(4\pi)^2}\left[\frac{1}{\epsilon}-\gamma+\ln\left(\frac{4\pi\mu^2}{M^2}\right)\right]
\int d^4x\, V_f^2 \cr
&\hspace{1cm}
-\frac{1}{2(4\pi)^2} \int \frac{d^4k}{(2\pi)^4}\,\widetilde{V}_f(k)\widetilde{V}_f(-k)
\int_0^1dx\,\ln\left[1-x(1-x)\frac{k^2}{M^2}\right]\,,
\label{eq:fb1}\end{align}
with the Fourier transform $\widetilde{V}(k)=\bigintsss d^Dx\, V(x){\rm e}^{\imu k_\nu x^\nu}$.
We define the normalization of $V_f(r)$ such that the $1/\epsilon$ in
singularities in Eq.~(\ref{eq:fb1}) match those of $C_g$. That is, for the given potential $V_f$ 
we scale all second order contributions by the factor $c_B$, which is determined from
\begin{equation}
\frac{-6c_B}{e^2}\int \frac{d^4x}{TL}\,V_f^2=
\int \frac{d^4x}{TL}\,F_{\mu\nu}F^{\mu\nu}
=4\pi v^2\int_0^\infty \rho d\rho\,\left(\frac{n^2g^{\prime2}}{\rho}\right)^2\,,
\label{eq:fb2}\end{equation}
where $TL$ is the volume of the subspace in which the vortex is translationally invariant.
Then we end up with the finite expression 
\begin{align}
C_g\int \frac{d^4x}{TL}\,F_{\mu\nu}F^{\mu\nu}+c_B\mathcal{A}^{\rm (fb)}_2&=
\frac{M^2}{288\pi}\left[22-5\sqrt{3}\pi\right]
\int_0^\infty \rho d\rho\left(n\frac{g^\prime}{\rho}\right)^2 \cr &
+c_B\frac{M^2}{16\pi}\int_0^\infty dq\,\widetilde{v}^2_f(q)
\left[\sqrt{q^2+8}\,{\rm asinh} \left(\frac{q}{\sqrt{8}}\right)-q\right]\,,
\label{eq:fb3}\end{align}
where
$\widetilde{v}_f(q)=\bigintsss_0^\infty \rho d\rho\,V_f(r)J_0(q\rho)$
and $J_0$ is a cylindrical Bessel function.

Above we have outlined the fake boson approach for $D=3+1$, where it is
needed to remove an ultra-violet divergence. Though it is not
required in $D=2+1$, we apply it there as well because the numerical
evaluation of the momentum space integrals in 
Sec.~\ref{sec:vpe} is more stable with the corresponding subtraction.
The correspondingly renormalized part of the action is
\begin{align}
C_g\int d^3x\,F_{\mu\nu}F^{\mu\nu}+c_B\mathcal{A}^{\rm (fb)}_2&=
-\frac{M}{96}\left[4+3\ln(3)\right]
\int_0^\infty \rho d\rho\left(n\frac{g^\prime}{\rho}\right)^2
-c_B\frac{M}{4\sqrt{2}}\int_0^\infty dq\,\widetilde{v}^2_f(q)
\arctan\left(\frac{q}{\sqrt{8}}\right) \,.
\label{eq:fb4}\end{align}

\section{Vacuum polarization energy (VPE)}
\label{sec:vpe}

With all the ingredients of the calculation determined, we next show
how to assemble them to determine the fully renormalized VPE.

\subsection{Relevance of scattering data}

In this Section we briefly review the spectral methods for computing the 
VPE of static, extended field configurations from Ref.~\cite{Graham:2009zz}.

The background field configuration induces a potential for small amplitude 
fluctuations, which are treated by standard techniques of scattering
theory in quantum mechanics. These calculations provide the bound
state energies, $\omega_j$,  which directly enter the VPE,
as well as the phase shifts $\delta(k)$, or more generally the
scattering matrix, as functions of the wave-number $k$ for
single-particle energies above the threshold given by the 
mass~$m$ of the fluctuating field. Those phase shifts parameterize the
change in the density of continuum modes via the Friedel-Krein
formula~\cite{Faulkner:1977aa},
\begin{equation}
\Delta\rho_\ell=\frac{1}{\pi}\frac{d\delta_\ell(k)}{dk}\,,
\label{eq:Krein}
\end{equation}
where $\ell$ indexes the partial wave-expansion in
Eq.~(\ref{eq:partialwave}). As shown in Eq.~(\ref{eq:VPE1}), that change 
determines the continuum contribution to the VPE
\begin{equation}
\Delta E=\frac{1}{2}\sum_{j}^{\rm b.s.}\omega_j
+\int_0^\infty \frac{dk}{2\pi}\sum_\ell
\sqrt{k^2+m^2}\,\frac{d\delta_\ell(k)}{dk}+E_{\rm CT}\,.
\label{eq:vpe1}\end{equation}
Our scattering problem is in two space dimensions and all angular
momentum sums run over the integers from negative to positive infinity. 

Next we describe how the counterterms cancel the divergences
originating from the large $k$ behavior of the phase shift in the
momentum integral. In the previous Section we have shown that the
Feynman diagrams are generated by expanding
the effective action in powers of the potential appearing in the
scattering wave-equations. The equivalent Born expansion for the phase
shifts is most efficiently performed within the variable phase approach
\cite{variable}, which factors out the outgoing wave from the
full wave-function. We will provide details of that approach for the
vortex problem  in the next Section and restrict ourselves here to the
description of the main concepts. The factor function is called the
Jost solution $\mathcal{F}_\ell(r,k)$, and the differential
equation for the Jost solution is (numerically) solved with the
boundary condition  $\lim_{r\to\infty}\mathcal{F}(r,k)=\ID$. In a
multi-channel problem $\mathcal{F}_\ell(r)$
is matrix-valued. Regularity of the scattering wave-function then
determines the scattering matrix and subsequently the phase shift for
a particular partial wave 
$$
\delta_\ell(k)=\frac{1}{2\imu}\ln \det \lim_{r\to0}
\left[\mathcal{F}_\ell^{\ast}(r,k)\mathcal{F}_\ell^{-1}(r,k)\right]\,.
$$
Most importantly, the Jost solution has a perturbation expansion in powers of the scattering
potential:
$\mathcal{F}_\ell(r,k)=\ID+\mathcal{F}_\ell^{(1)}(r,k)+
\mathcal{F}_\ell^{(2)}(r,k)+\ldots$ 
with boundary conditions $\lim_{r\to\infty}\mathcal{F}_\ell^{(n)}(r,k)=0$. The individual contributions
can be straightforwardly obtained by iterating the wave-equation. This
expansion in turn induces the Born series for the phase shift,
\begin{align}
\delta_\ell^{(1)}(k)&=\frac{1}{2\imu}\lim_{r\to0}\,{\rm tr} 
\left[\mathcal{F}_\ell^{(1)\ast}(r,k)-\mathcal{F}_\ell^{(1)}(r,k)\right]\,,\cr
\delta_\ell^{(2)}(k)&=\frac{1}{2\imu}\lim_{r\to0}\,{\rm tr}
\left[\mathcal{F}_\ell^{(2)*}(r,k)-\mathcal{F}_\ell^{(2)}(r,k)
-\frac{1}{2}[\mathcal{F}_\ell^{(1)}(r,k)]^2+\frac{1}{2}[\mathcal{F}_\ell^{(1)*}(r,k)]^2\right]\,,
\qquad \mbox{etc.}
\label{eq:deltaBorn}
\end{align}
Since the phase shift is dimensionless, the expansion in powers of the potential is also an 
expansion in the inverse momentum. Hence taking sufficiently many terms, $N$, from the Born series 
and subtracting them from the phase shift will render the momentum integral in Eq.~(\ref{eq:vpe1}) 
finite. We then add these subtractions back via the equivalent
Feynman expansion up to order $N$ that we developed in the previous Section,
\begin{equation}
\Delta E=\frac{1}{2}\sum_{j}^{\rm b.s.}\omega_j
+\int_0^\infty \frac{dk}{2\pi}\sum_\ell \sqrt{k^2+m^2}\,
\frac{d}{dk}\left[\delta_\ell(k)-\delta_\ell^{(1)}(k)-\delta_\ell^{(2)}(k)-\ldots-\delta_\ell^{(N)}(k)\right]
+E_{\rm FD}^{(N)}+E_{\rm CT}\,.
\label{eq:vpe2}\end{equation}
The sum $E_{\rm FD}^{(N)}+E_{\rm CT}$ combines to form an
ultra-violet finite expression as, for example,
for the action in Eqs.~(\ref{eq:A2D3}) and (\ref{eq:A2D4}).
We note that this process is exact within one loop and does not rely on
the accuracy of the Born approximation.

To further process the momentum integral we recall that the Jost function\footnote{The general definition 
of the Jost function is the Wronskian of the Jost and regular solutions.} 
$\mathcal{F}(k)=\lim_{r\to0}\mathcal{F}(r,k)$ is analytic for ${\sf Im}(k)\ge0$ and that for real $k$ its 
complex conjugate is $\mathcal{F}^\ast(k)=\mathcal{F}(-k)$ \cite{Newton:1982qc}. Thus we write
$\delta_\ell(k)=(1/2\imu)\left[\ln \det
\mathcal{F}_\ell(-k)- \ln \det \mathcal{F}_\ell(k)\right]$ 
and extend the integral over the full real axis. With the Born
subtractions made above, there will be no
contribution to the integral from a semi-circle at infinitely large
$|k|$ for ${\sf Im}(k)\ge0$, and we may
thus close the contour accordingly. However, we have to circumvent the branch cut that emerges in 
$\sqrt{k^2+m^2}$ for ${\sf Im}(k)>m$. That will leave a contribution along the imaginary $k$ axis starting 
at $\imu m$ that picks up the discontinuity of the square root. Finally, we collect the residues that
emerge from the zeros of the Jost function, which create first-order poles in the logarithmic derivative. 
These zeros are known to be single and located at the wave-numbers corresponding to the bound state energies: $k_j=\imu\sqrt{m^2-\omega_j^2}$
\cite{Newton:1982qc}. By virtue of Cauchy's theorem, these residues exactly 
cancel the discrete bound state contribution in the VPE
\cite{Bordag:1996fv}. Introducing $t=k/{\imu}$ and 
$\nu_\ell(t)= \ln \det \mathcal{F}_\ell(\imu t)$
and integrating by parts, we obtain the compact expression
\begin{equation}
\Delta E=\int_m^\infty \frac{dt}{2\pi} \frac{t}{\sqrt{t^2-m^2}}\sum_\ell
\left[\nu_\ell(t)-\nu^{(1)}_\ell(t)-\nu^{(2)}_\ell(t)-\ldots-\nu^{(N)}_\ell(t)\right]
+E_{\rm FD}^{(N)}+E_{\rm CT}\,,
\label{eq:vpe3}\end{equation}
where the $\nu^{(n)}_\ell(t)$ arise from the Born series expansion of $\mathcal{F}_\ell(\imu t)$.

The situation in $D=3+1$ is slightly more complicated. The vortex is translationally invariant
along the symmetry axis (which we choose to be in the $z$ direction). The wave-function has 
a plane wave factor for the dependence of that coordinate and we have to integrate over the corresponding  
momentum. Since the phase shifts do not depend on that momentum, there is no Born subtraction that removes 
the ultra-violet divergence emerging from that integration. The solution is the so-called interface 
formalism developed in Ref.~\cite{Graham:2001dy}, in which that integral is dimensionally regularized in 
$d-1$ space dimensions, showing that the divergence is proportional to 
$$
\frac{1}{d-1}\left\{\int_0^\infty \frac{dk}{\pi} k^2 \frac{d}{dk}
\sum_\ell \left[\delta_\ell(k)-\delta_\ell^{(1)}(k)-\delta_\ell^{(2)}(k)-\ldots-\delta_\ell^{(N)}(k)\right]
+\sum_j^{\rm b.s.} \omega_j^2-m^2\right\}\,.
$$
The expression in curly brackets actually vanishes in all partial wave channels individually via one 
of the sum rules that generalize Levinson's theorem \cite{Graham:2001iv}, which follow from analyticity 
of the Jost function for ${\sf Im}(k)\ge0$. Thus the limit $d\to1$ can be taken, yielding
\begin{equation}
\Delta E = \frac{-1}{8\pi} \left[\int_0^\infty \frac{dk}{\pi}\sum_\ell \omega^2(k)
\ln\frac{\omega^2(k)}{\overline{\Lambda}^2}\, \frac{d}{dk}
\left[\nu_\ell(t)-\nu^{(1)}_\ell(t)-\nu^{(2)}_\ell(t)-\ldots-\nu^{(N)}_\ell(t)\right]
+ \sum_j^{\rm b.s.}
\omega_{j}^2\ln\frac{\omega_{j}^2}{\overline{\Lambda}^2}\right] +E^{(N)}_{\rm FD}+E_{\rm CT}
\label{eq:inter1}
\end{equation}
as the VPE per unit length of the vortex. The arbitrary energy scale
$\overline{\Lambda}$ has been introduced for dimensional reasons. It
cancels in Eq.~(\ref{eq:inter1}) by a generalization of Levinson's
theorem. Again, by closing the contour in the upper half $k$-plane, we
can remove the  explicit bound state contribution. This time we pick
up the discontinuity from the logarithm,
\begin{equation}
\Delta E =\int_m^\infty\frac{dt}{4 \pi}\, t\, 
\sum_\ell\left[\nu_\ell(t)-\nu^{(1)}_\ell(t)-\nu^{(2)}_\ell(t)-\ldots-\nu^{(N)}_\ell(t)\right]
+E^{(N)}_{\rm FD}+E_{\rm CT}\,.
\label{eq:inter2}
\end{equation}

\subsection{Jost function for scattering about the vortex}

Because the singularity of the vortex field configuration
at its center makes it impossible to straightforwardly apply the
standard form of the spectral methods described above. In this Section
we explain the required modifications. As described above,
the ultra-violet singularities that occur at third
and higher order in the expansion of the effective action cancel (when
regularized in a gauge invariant scheme), so we may set $N=2$ hereafter.

We return to the dimensionless variables $\rho=evr$ and $q=k/ev$ that
enter the wave-equations~(\ref{eq:fluct1}). We define the Jost
solution by introducing
$\Psi_\ell\sim\left(\begin{smallmatrix}\eta_\ell \cr a_{\ell+1}\end{smallmatrix}\right)$ 
in scattering channel $\ell$ as a $2\times2$ matrix with the free
solution factored out,
\begin{equation}
\Psi_\ell=\mathcal{F}_\ell\cdot\mathcal{H}_\ell
\qquad{\rm where}\qquad
\mathcal{H}_\ell=\begin{pmatrix}H^{(1)}_{\ell}(q\rho)& 0
\cr 0& H^{(1)}_{\ell+1}(q\rho)\end{pmatrix}\,,
\label{eq:param}\end{equation}
with boundary condition $\lim_{\rho\to\infty}\mathcal{F}_\ell=\ID$. Since the two columns
of $\mathcal{H}_\ell$ represent free outgoing cylindrical waves for either $\eta_\ell$ or $a_{\ell+1}$,
the physical scattering solution is
$$
\Psi^{\rm sc}_\ell=\mathcal{F}^\ast_\ell\cdot\mathcal{H}^\ast_\ell-
\mathcal{F}_\ell\cdot\mathcal{H}_\ell\cdot\mathcal{S}_\ell\,,
$$
where $\mathcal{S}_\ell$ is the scattering matrix, which can extracted using
$\lim_{\rho\to0}\Psi_{\rm sc}=0$. Then
$\mathcal{F}_\ell=\mathcal{F}_\ell(\rho,q)$ is the Jost solution,
which leads to the Jost function
$\mathcal{F}_\ell(q)=\lim_{\rho\to0}\mathcal{F}_\ell(\rho,q)$. In both
cases, $D=2+1$ and $D=3+1$, a major ingredient for the scattering
piece of the VPE is the Jost function for imaginary momenta
$\nu_\ell(t)= \ln \det\left[\mathcal{F}_\ell(\imu t)\right]$. In
matrix form, the scattering differential equations read
\begin{equation}
\frac{\partial^2}{\partial\rho^2}\mathcal{F}_\ell
=-\frac{\partial}{\partial\rho}\mathcal{F}_\ell
-2\left(\frac{\partial}{\partial\rho}\mathcal{F}_\ell\right)\cdot\mathcal{Z}_\ell
+\frac{1}{\rho^2}\left[\mathcal{L}_\ell,\mathcal{F}_\ell\right]
+\mathcal{V}_\ell\cdot\mathcal{F}_\ell\,.
\label{eq:jostdeq}\end{equation}
The angular momenta enter via the derivative matrix for the
analytically continued Bessel functions
\begin{equation}
\mathcal{Z}_\ell=\begin{pmatrix}
\frac{|l|}{\rho}-t\,\frac{K_{|l|+1}(t\rho)}{K_{|l|}(t\rho)} & 0 \cr
0 & \frac{|l+1|}{\rho}-t\,\frac{K_{|l+1|+1}(t\rho)}{K_{|l+1|}(t\rho)}
\end{pmatrix}
\qquad {\rm and}\qquad
\mathcal{L}_\ell=\begin{pmatrix}\ell^2 & 0 \cr 
0 & (\ell+1)^2 \end{pmatrix}\,,
\label{eq:angmom}\end{equation}
and the potential matrix is
\begin{equation}
\mathcal{V}_\ell=\begin{pmatrix}
3(h^2(\rho)-1)+\frac{1}{\rho^2}(n^2g^2(\rho)-2n\ell g(\rho))\qquad & \sqrt{2}d(\rho)\cr
\sqrt{2}d(\rho)& 2(h^2(\rho)-1) \end{pmatrix}\,.
\label{eq:potmat}\end{equation}
The standard procedure to determine the Born approximations, which are
needed to regularize the ultraviolet divergences, fails when $g(0)\ne0$ 
\cite{Graham:2019fzo}. This can be seen by noting that the integral 
$\int_0^\infty \rho d\rho\,\left(\frac{g(\rho)}{\rho}\right)^2$, which
would appear in the leading Born approximation for the gauge field potential,
is ill-defined in the singular gauge.  To perform the Born subtractions without 
the singular terms, we introduce
\begin{equation}
\overline{\mathcal{V}}=\begin{pmatrix}
3(h^2(\rho)-1) & \sqrt{2}d(\rho)\cr
\sqrt{2}d(\rho)& 2(h^2(\rho)-1) \end{pmatrix}
\label{eq:potmat1}\end{equation}
and iterate the auxiliary differential equation
\begin{equation}
\frac{\partial^2}{\partial\rho^2}\overline{\mathcal{F}}_\ell=-\frac{\partial}{\partial\rho}\overline{\mathcal{F}}_\ell
-2\left(\frac{\partial}{\partial\rho}\overline{\mathcal{F}}_\ell\right)\cdot\mathcal{Z}_\ell
+\frac{1}{\rho^2}\left[\mathcal{L}_\ell,\overline{\mathcal{F}}_\ell\right]
+\overline{\mathcal{V}}\cdot\overline{\mathcal{F}}_\ell
\label{eq:barjostdeq}\end{equation}
according to the expansion
$\overline{\mathcal{F}}_\ell=\ID+\overline{\mathcal{F}}_\ell^{(1)}+\overline{\mathcal{F}}_\ell^{(2)}+\ldots$. 
The relevant leading orders are
\begin{align}
\frac{\partial^2}{\partial\rho^2}\overline{\mathcal{F}}_\ell^{(1)}
&=-\frac{\partial}{\partial\rho}\overline{\mathcal{F}}_\ell^{(1)}
-2\left(\frac{\partial}{\partial\rho}\overline{\mathcal{F}}_\ell^{(1)}\right)\cdot\mathcal{Z}_\ell
+\frac{1}{\rho^2}\left[\mathcal{L}_\ell,\overline{\mathcal{F}}_\ell^{(1)}\right]
+\overline{\mathcal{V}}\,,\cr
\frac{\partial^2}{\partial\rho^2}\overline{\mathcal{F}}_\ell^{(2)}
&=-\frac{\partial}{\partial\rho}\overline{\mathcal{F}}_\ell^{(2)}
-2\left(\frac{\partial}{\partial\rho}\overline{\mathcal{F}}_\ell^{(2)}\right)\cdot\mathcal{Z}_\ell
+\frac{1}{\rho^2}\left[\mathcal{L}_\ell,\overline{\mathcal{F}}_\ell^{(2)}\right]
+\overline{\mathcal{V}}\cdot\overline{\mathcal{F}}_\ell^{(1)}\,,
\label{eq:Born1}\end{align}
and all $\overline{\mathcal{F}}_\ell^{(m)}$ vanish in the limit
$\rho\to\infty$. From the differential equations~(\ref{eq:jostdeq}) and~(\ref{eq:Born1}) we extract
\begin{equation}
\nu_\ell(t)=\lim_{\rho\to\rho_{\rm min}} \ln \det
\mathcal{F}_\ell\,,\quad
\overline{\nu}_\ell^{(1)}(t)=\lim_{\rho\to\rho_{\rm min}}{\rm tr}\overline{\mathcal{F}}_\ell^{(1)}\quad {\rm and}\quad 
\overline{\nu}_\ell^{(2)}(t)=\lim_{\rho\to\rho_{\rm min}}{\rm tr}\left[\overline{\mathcal{F}}_\ell^{(2)}
-\frac{1}{2}\left(\overline{\mathcal{F}}_\ell^{(1)}\right)^2\right]
\label{eq:compnu}
\end{equation}
where $\rho_{\rm min}$ is a tiny but nonzero number. The above
expansion is the analog of Eq.~(\ref{eq:deltaBorn}) for the Jost
function of the vortex with imaginary momentum, but our Born
subtraction no longer includes the singular terms; we describe how to
handle them below.

The Jost functions for the ghost and fake boson are analogous to the above, but much simpler.
The Jost solution is no longer a matrix, so the commutator term disappears. Then one 
just replaces $\mathcal{V}$ and $\overline{\mathcal{V}}$ by $2v^2\left[h^2(\rho)-1\right]$
for the ghost and by $V_f(\rho)$ for the fake boson. Furthermore, the
angular momentum sum is symmetric in $\ell \to -\ell$, so it can be
simplified to run over non-negative values with a degeneracy factor of
two for $\ell\ge1$.

By subtracting just $\overline{\nu}_\ell^{(1)}(t)$ and
$\overline{\nu}_\ell^{(2)}(t)$ from 
$\nu_\ell(t)$ we do not include the subtractions for the singular
terms that, in a gauge-invariant formulation, induce a logarithmic
ultra-violet divergence for $D=3+1$. To investigate the relevant diagrams we
compare the dimensional and sharp cut-off regularization schemes, leading to
the identification (with an arbitrary mass scale $\Lambda$),
\begin{equation}
\frac{1}{\epsilon(4\pi)^2}=-\imu\int\frac{d^4l}{(2\pi)^4}\,
\frac{1}{\left(l^2-\Lambda^2+\imu0^+\right)^2}\Bigg|_{\rm div.}
=\frac{1}{8\pi^2}\int \frac{l^2 dl}{\sqrt{l^2+\Lambda^2}^3}\Bigg|_{\rm div.}\,,
\end{equation}
We thus expect for $\rho_{\rm min}\to0$
\begin{equation}
\left[\nu(t)\right]_V:=
\lim_{L\to\infty}\sum_{\ell=-L}^{L}\left[\nu_\ell(t)-\overline{\nu}^{(1)}_\ell(t)
-\overline{\nu}^{(2)}_\ell(t)\right]_{\rho_{\rm min}}
-n^2\int_{\rho_{\rm min}}^\infty \frac{d\rho}{\rho}\,g^2(\rho)
\,\stackrel{t\to\infty}{\longrightarrow}\,
\frac{n^2}{12t^2}\int_0^\infty \frac{d\rho}{\rho}\,\left(\frac{dg(\rho)}{d\rho}\right)^2\,.
\label{eq:sing1}\end{equation}
Again, the superscripts denote the Born expansion order with respect to $\overline{\mathcal{V}}$.
As explained in Ref.~\cite{Graham:2019fzo}, the integral subtraction
on the left-hand side relates to a quadratic divergence in the VPE. By
this subtraction we restore gauge invariance, which is not manifest
for the Jost function. Note that we can write that integral as
$$
n^2\int_{\rho_{\rm min}}^\infty \frac{d\rho}{\rho}\,g^2(\rho)
=\sum_\ell \int_{\rho_{\rm min}}^\infty \frac{d\rho}{\rho}\, J_\ell^2(qr)
\left[n^2g^2(\rho)-2n\ell g(\rho)\right]\,,
$$
where the $J_\ell(z)$ are cylindrical Bessel functions. This integral then
replaces the leading Born approximation from the singular terms in the wave-equation.
Its contribution to $\nu_\ell(t)$ arises from an integration by parts of 
an expression that contains its derivative with respect to $t$, {\it
cf.} Eq.~(\ref{eq:vpe1}).  Hence subtracting a constant times this quantity
has no effect on the result, but renders the integral well-defined on
the imaginary axis.

In $D=3+1$, the right-hand side of Eq.~(\ref{eq:sing1}) still contains
a logarithmic divergence, which we computed in dimensional
regularization. It actually arises from a combination of two
Feynman diagrams that individually are quadratically divergent, {\it cf.} Eq.~(\ref{eq:AeffA}). 
The analogous Born expansions would have to be performed individually. However, for 
the singular vortex configuration, these integrals do not exist. Hence we apply the fake boson 
formalism developed in Ref.~\cite{Farhi:2001kh} as described above. As shown in 
Eq.~(\ref{eq:fb3}), the second-order term for a scalar field also yields a logarithmic 
divergence.  In principle, the strength of that divergence does not depend on the mass of 
the boson. We take it to equal the classical Higgs/gauge mass so that we can simply subtract 
the associated Born term from $\nu_\ell(t)$ with a suitably adjusted strength and add it back 
as a Feynman diagram. To be precise, we consider scattering of a boson in the potential 
$V_f(\rho)=3(\tanh^2(\kappa \rho)-1)$, for which $\overline{\nu}^{(2)}_\ell(t)$ is the second-order 
contribution to the Jost function on the imaginary momentum axis. The partial wave expansion for this 
scalar field is similar to that of ghost field in Eq.~(\ref{eq:fluc2}) with $2(h(\rho)-1)$ replaced 
by $V_f(\rho)$. We take $\kappa$ as an arbitrary parameter to later test our numerical simulation, 
since the final result for VPE should not depend on a particular choice for $V_f$. This subtraction 
is calibrated by Eq.~(\ref{eq:fb2}), which for this particular scalar potential reads
\begin{equation}
c_B=-\frac{e^2}{6}\frac{\int_0^\infty rdr\, F_{\mu\nu}F^{\mu\nu}}
{\int_0^\infty r dr V_f^2}
=-\frac{n^2}{3}\frac{\int_0^\infty \rho d\rho
\left(\frac{g^\prime(\rho)}{\rho}\right)^2}
{\int_0^\infty \rho d\rho\, \left[3(\tanh^2(\kappa \rho)-1)\right]^2}\,.
\label{eq:CB}\end{equation}

\subsection{VPE for $D=2+1$ and $D=3+1$}

As mentioned above, our analysis proceeds with dimensionless variables
such that $ev=1$. We return to dimensionful expressions by multiplying
with appropriate powers of $\frac{M}{\sqrt{2}}$ where $M$ is the Higgs
mass, which does not acquire quantum correction in our on-shell
scheme.

For $D=3+1$, the ghost and non-transverse gauge field contributions
cancel. In that case, however, we still have to integrate over the
momentum conjugate to the symmetry axis using the interface formalism
of Eq.~(\ref{eq:inter2}). For later discussion we separate
the scattering contribution (including the factor of two for the
complex fields) after the fake boson subtraction,
\begin{equation}
\Delta E_{\rm scat.}=\frac{M^2}{2}\int_{\sqrt{2}}^\infty \frac{dt}{2\pi}t
\left\{\left[\nu(t)\right]_V-c_B \nu_{\rm B}(t)\right\}\,.
\label{eq:Escat3}
\end{equation}
Here $\nu_{\rm B}$(t) is the angular momentum sum of the second Born contribution 
to the logarithm of the Jost function from the fake boson potential.

To evaluate the renormalized Feynman diagram contributions, we
introduce Fourier transforms of the vortex profiles
\begin{equation}
\begin{array}{lcl}
I_A(q)=n\bigintsss_0^\infty d\rho\, h(\rho) g(\rho) J_1(q\rho)\,,\quad
&\qquad\qquad
&I_H(q)=q\bigintsss_0^\infty \rho d\rho\,\left[1-h(\rho)\right]J_0(q\rho)\,,\\[3mm]
\widetilde{v}_H(q)=\bigintsss_0^\infty \rho d\rho\,\left[h^2(\rho)-1\right]J_0(q\rho)\,,\quad
&&\widetilde{v}_f(q)=3\bigintsss_0^\infty \rho d\rho\,\left[\tanh^2(\kappa\rho)-1\right]J_0(q\rho)\,,
\label{eq:Ftrans1}
\end{array}
\end{equation}
where, again, the $J_\ell(z)$ are cylindrical Bessel functions.

In $D=3+1$ we have logarithmic divergences at second order, and we
therefore separate the finite counterterm
\begin{align}
E_{\rm CT}&=\frac{M^2}{16\pi}\int_0^\infty \rho d\rho\left\{
\left[17-10\frac{\pi}{\sqrt{3}}\right]\left[h^{\prime2}+n^2\frac{h^2}{\rho^2}g^2\right]
+\frac{1}{2}\left[35-19\frac{\pi}{\sqrt{3}}\right]\left[h^2-1\right]^2
+\frac{c_B}{18}\left[22-5\sqrt{3}\pi\right]\left(\frac{ng^\prime}{\rho}\right)^2\right\}\,.
\label{eq:ECT4}
\end{align}
and the finite Feynman diagram contributions
\begin{align}
E_{\rm FD}&=\frac{M^2}{2\pi}\int_0^\infty dq \left[I_A^2(q)+I_H^2(q)+\frac{13}{8}\widetilde{v}_H^2(q)
+\frac{c_B}{8}\widetilde{v}_f^2(q)\right]
\left[\sqrt{q^2+8}\,{\rm asinh}\left(\frac{q}{\sqrt{8}}\right)-q\right]\,.
\label{eq:EFD4} \end{align}
Recall that $\widetilde{v}_f(q)$ was defined after Eq.~(\ref{eq:fb3}).

In $D=2+1$, the second order contributions do not lead to an ultra-violet divergence.
Nevertheless it is convenient to subtract them from the scattering data and add them back 
as Feynman diagrams, because it allows us to use the same $\left[\nu(t)\right]_{\rm V}$ 
as above. We separate the scattering contribution in
Eq.~(\ref{eq:vpe3}) and augment it by the ghost piece
\begin{equation}
\Delta E_{\rm scat.}=\frac{M}{\sqrt{2}}\left\{\int_{\sqrt{2}}^\infty \frac{dt}{\pi}
\frac{t}{\sqrt{t^2-2}}\,\left\{\left[\nu(t)\right]_V-c_B \nu_{\rm B}(t)\right\}
-\int_{\sqrt{2}}^\infty \frac{dt}{2\pi} \frac{t}{\sqrt{t^2-2}}\,\nu_{\rm gh}(t)\right\}\,,
\label{eq:Escat2}
\end{equation}
where $\nu_{\rm gh}(t)$ is the angular momentum sum of the logarithm of the Jost function with two
Born subtractions for the single channel potential $V_{\rm gh}=2(\Phi^2_S-v^2)=2v^2(h^2(\rho)-1)$.
In comparison with Eq.~(\ref{eq:vpe3}), a factor of two again appears in the first integral because we are dealing
with a complex boson field. Note that this first integral in Eq.~(\ref{eq:Escat2}) would also be 
finite without the fake boson subtraction. However, its inclusion
improves the large momentum convergence 
of that integral and is thus advantageous in the numerical simulation.

The final ingredient is the renormalized Feynman diagram contribution in $D=2+1$,
\begin{align}
E_{\rm FD}&=-\sqrt{2}M\int_0^\infty dq \left[I_A^2(q)+I_H^2(q)+\frac{11}{8}\widetilde{v}_H^2(q)
+\frac{c_B}{8}\widetilde{v}_f^2(q)\right]
\arctan\left(\frac{q}{\sqrt{8}}\right)\cr
E_{\rm CT}&=\frac{M}{48}\int_0^\infty \rho d\rho\left\{
\left[45\ln(3)-28\right]\left[h^{\prime2}+n^2\frac{h^2}{\rho^2}g^2\right]
+\frac{1}{2}\left[87\ln(3)-28\right]\left[h^2-1\right]^2
+\frac{1}{2}\left[3\ln(3)+4\right]\left(\frac{ng^\prime}{\rho}\right)^2\right\}\,.
\label{eq:EFD3}
\end{align}
As in Eq.~(\ref{eq:EFD4}), factors of $1/\sqrt{2}$ emerged in the
arguments of the trigonometric and hyperbolic functions in the
counterterm contribution because $M=\sqrt{2}ev$.

\section{Numerical results for the VPE}

As a first step we substitute the profile functions with the
parameterization of 
Eq.~(\ref{eq:profilefit}) into $\mathcal{V}$ and $\overline{\mathcal{V}}$ in 
Eqs.~(\ref{eq:potmat}) and~(\ref{eq:barjostdeq}), respectively. 
For a given angular momentum channel, we then integrate 
the differential equations~(\ref{eq:jostdeq}) and~(\ref{eq:Born1})
with the appropriate  boundary conditions from a large 
$\rho_{\rm  max}\approx20$ to a small $\rho_{\rm min}$ near the center
of the vortex. Once $\rho_{\rm min}$ is small enough, we compute
\begin{equation}
\nu_L(t)=\sum_{\ell=-L}^{L}\left[\nu_\ell(t)-\overline{\nu}^{(1)}_\ell(t)
-\overline{\nu}^{(2)}_\ell(t)\right]_{\rho_{\rm min}}
-n^2\int_{\rho_{\rm min}}^\infty \frac{d\rho}{\rho}\,g^2(\rho)\,.
\label{eq:largeL}\end{equation}
We find that there are still small variations when decreasing 
$\rho_{\rm min}$ even further. These variations emerge from channels 
that do not have an angular momentum barrier at the center of the vortex. 
In those instances, the regular and irregular solutions respectively
approach a constant and logarithm, which are difficult to disentangle
numerically. According to Eq.~(\ref{eq:fluct1}) these are the
$\ell=-1$ and $\ell=n$ channels. In those channels we use several
small values for $\rho_{\rm min}$ and fit
$$
\ln \det \mathcal{F}_\ell-\overline{\mathcal{F}}_\ell^{(1)}
-\overline{\mathcal{F}}_\ell^{(2)}
-\frac{1}{2}\left(\overline{\mathcal{F}}_\ell^{(1)}\right)^2
=a_0 +\frac{a_1}{\ln (\rho_{\rm min})}
+\frac{a_2}{\ln^2(\rho_{\rm min})}
$$
and replace the square bracket in Eq.~(\ref{eq:largeL}) by $a_0$ for these 
$\ell$-values. We note that this fit is also needed for regular gauge profiles 
even though they have well-defined Born series, because there one
still has to disentangle constant and logarithm behaviors \cite{Graham:2011fw}.
We remark that the cancellations of the $\rho_{\rm min}$ singularities in 
Eq.~(\ref{eq:largeL}) stems from the large $L$ terms in that sum.

\begin{figure}
\centerline{\epsfig{file=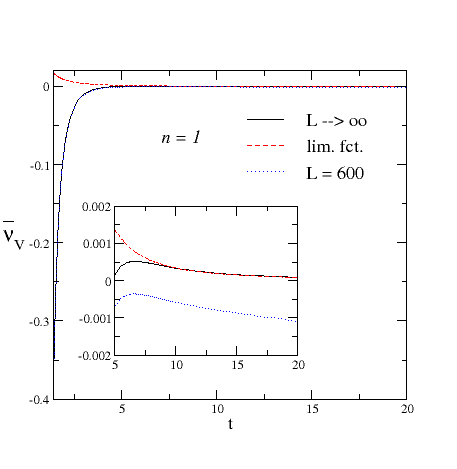,width=7cm,height=4cm}~~~
\epsfig{file=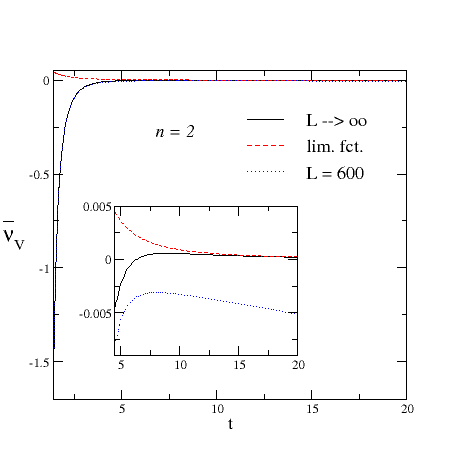,width=7cm,height=4cm}}
\centerline{\epsfig{file=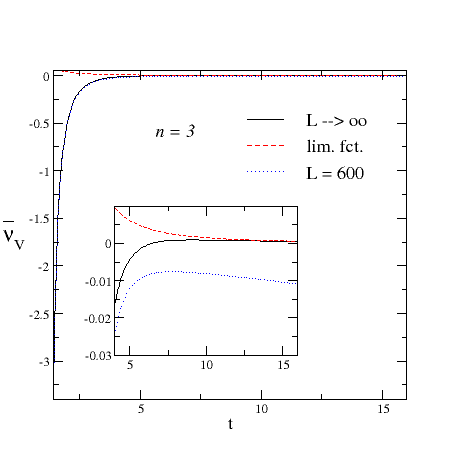,width=7cm,height=4cm}~~~
\epsfig{file=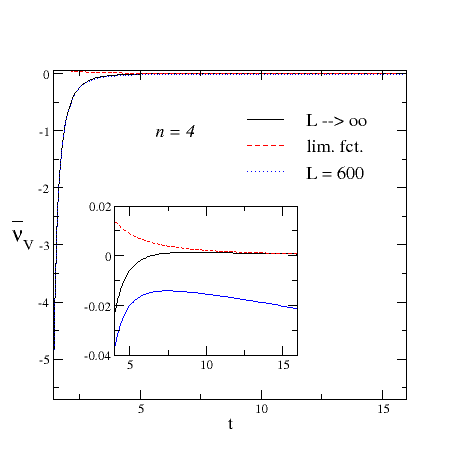,width=7cm,height=4cm}}
\caption{\label{fig:largeL}Large angular momentum extrapolation from Eq.~(\ref{eq:largeL}). 
The line labeled {\it lim.\ fct.} refers to the large $t$ behavior in
Eq.~(\ref{eq:sing1}).}
\end{figure}

It is essential to verify that $\left[\nu(t)\right]_V$
exhibits the asymptotic behavior predicted in Eq.~(\ref{eq:sing1}) by
the analysis of the two-point function for the gauge field.  It turns
out that even summing up to a large value like $L=600$ is insufficient
to compute the sum on the left-hand side. As explained in
Ref.~\cite{Graham:2019fzo}, on top of computing the sum for such large
values of $L$, an extrapolation for $L\to\infty$ is needed. This is
done by using different large values of $L$ in Eq.~(\ref{eq:largeL}) 
and extracting $\left[\nu(t)\right]_V$ from a fit of the form
$\nu_L(t)\approx\left[\nu(t)\right]_V+\frac{b_1}{L}+\frac{b_2}{L^2}$.
The importance of this extrapolation is shown in the inserts of
Fig.~\ref{fig:largeL}. Though the numerical effect appears to be
small, we note that the integrand for $D=3+1$ in Eq.~(\ref{eq:Escat3}) 
has an additional factor of $t$, which amplifies any inaccuracy at
large $t$. Also, without that extrapolation the integrand may incorrectly 
appear to converge without subtractions already at moderate $t$
\cite{Graham:2019fzo}, which has led to incorrect conclusions
in the past \cite{Pasipoularides:2000gg}.

As suggested by this discussion, the numerical simulation is quite costly in computation 
time.\footnote{For large angular momenta sufficient accuracy can only be accomplished with 
{\it long-double} precision. This adds considerably to the computation time.} We therefore 
compute $\left[\nu(t)\right]_V$ for about 200 different $t$ values and implement a Laguerre 
interpolation to obtain a smooth function. This interpolation also allows for the substitution
to $\tau=\sqrt{t^2-2}$, which avoids the integrable singularity in Eq.~(\ref{eq:Escat2}).
The final VPE is simply the sum
\begin{equation}
\Delta E=E_{\rm FD}+E_{\rm CT}+\Delta E_{\rm scat.}\,.
\label{eq:totalVPE}\end{equation}

In Table~\ref{tab:D2} we present our results for the VPE of the vortex for $D=2+1$. 
We also list the Feynman diagram and counterterm contributions from Eq.~(\ref{eq:EFD3}) as well
as the scattering piece from Eq.~(\ref{eq:Escat2}) with the
extrapolations described above.
While the cancellations between the Feynman diagram and counterterm pieces are expected,
the fact that their combination goes in the same direction as the scattering piece 
is somewhat surprising. Overall, we find that the total VPE is always negative.
\begin{table}[h]
\centerline{
\begin{tabular}{c||c|c|c|c}
& $n=1$ & $n=2$ & $n=3$ & $n=4$ \cr
\hline
$E_{\rm CT}$        & 0.2671 & 0.4819 & 0.6786 & 0.8662 \cr
$E_{\rm FD}$        &-0.5156 &-1.1365 &-1.7588 &-2.3877 \cr
\hline
                          &-0.2484 &-0.6546 &-1.0801 &-1.5215 \cr
$\Delta E_{\rm scat.}$    &-0.0882 &-0.3408 &-0.6631 &-1.0205 \cr
\hline
$\Delta E$             &-0.3367 &-0.9955 &-1.7432 &-2.5420
\end{tabular}}
\caption{\label{tab:D2}Various contributions to the VPE of BPS vortices for $D=2+1$.
$E_{\rm CT}$ and $E_{\rm FD}$ are the sums of all listed counterterm contributions, including the 
fake boson. All data are in units of $M=\sqrt{2}ev$.}
\end{table}

\begin{table}[ht]
\centerline{
\begin{tabular}{c||c|c|c|c}
& $n=1$ & $n=2$ & $n=3$ & $n=4$ \cr
\hline
$E_{\rm CT}$        &-0.0134 &-0.0212 &-0.0272 &-0.0325 \cr
$E_{\rm FD}$        & 0.0212 & 0.0157 & 0.0158 & 0.0167 \cr
\hline
                          & 0.0078 &-0.0054 &-0.0114 &-0.0157 \cr
$\Delta E_{\rm scat.}$    &-0.0255 &-0.0969 &-0.1782 &-0.2622 \cr
\hline
$\Delta E$             &-0.0177 &-0.1023 &-0.1896 &-0.2784
\end{tabular}}
\caption{\label{tab:D3}Various contributions to the VPE of BPS vortices for $D=3+1$.
Scattering data as before, $E_{\rm CT}$ and $E_{\rm FD}$ are the sums of all listed 
counterterm contributions, including the fake boson. All data are in units of $M^2=2(ev)^2$.}
\end{table}

Table \ref{tab:D3} contains our results for the VPE per unit length of the 
vortex in $D=3+1$. The ingredients for Eq.~(\ref{eq:totalVPE}) are taken from 
Eqs.~(\ref{eq:Escat3}), (\ref{eq:ECT4}) and (\ref{eq:EFD4}).
Somewhat unexpectedly, the counterterm and Feynman diagram
contributions go in the same direction, presumably because $E_{\rm CT}$ 
involves timelike momenta while $E_{\rm FD}$ is an integral over spacelike
momenta. On the other hand there is a substantial cancellation between the scattering 
and Feynman diagram contributions once they are combined with the counterterms 
that implement the on-shell renormalization.
In both $D=2+1$ and $D=3+1$, $\Delta E_{\rm scat.}$ is sizable, which
is a clear  indication that the VPE cannot be reliably computed from
only low order Feynman diagrams.

In $D=3+1$ the VPE is essentially linear in the winding number $n$. We fit the data
from Table \ref{tab:D3} as: $\Delta E=-0.0166-0.0869(n-1)$. 
The quality of the fit is measured as $\chi^2=4.4\times10^{-6}$. To obtain a similarly 
small $\chi^2=7.2\times10^{-5}$ in $D=2+1$ we need to add a quadratic contribution 
$\Delta E=-0.3348-0.6314(n-1)-0.0350(n-1)^2$, but the coefficient of that
contribution is quite small.

We observe qualitatively similar winding number dependences in $D=2+1$ and $D=3+1$. An 
analogous similarity between these two cases was also found for the fermion VPE of QED flux 
tubes once equivalent renormalization conditions were implemented~\cite{Graham:2004jb}.

We have already performed a consistency test of our results when comparing the large 
momentum behavior of $\overline{\nu}_{\rm V}$ in Figure \ref{fig:largeL}. We have also verified 
the independence with respect to the fake boson potential $V_f(\rho)$ by using different
values for the mass parameter $\kappa$. We show an example in Table \ref{tab:fb} with two
$\kappa$ values for $n=4$ for both $D=2+1$ and $D=3+1$.
The entries for 
$\kappa=0.7$ are those from Tables \ref{tab:D2} and \ref{tab:D3}.
The variation with $\kappa$ of the individual components is small but
significant, and when combined to $\Delta E$ these variations indeed cancel.

\begin{table}[ht]
\parbox[l]{7cm}{
\begin{tabular}{c||c|c}
& $\kappa=0.7$ & $\kappa=1.0$ \cr
\hline
$\Delta E_{\rm scat.}$     &-1.02049 &-1.01919\cr
$E_{\rm FD}+E_{\rm CT}$    &-1.52148 &-1.52278\cr
$\Delta E$                 &-2.54197 &-2.54197
\end{tabular}}
\parbox[r]{7cm}{
\begin{tabular}{c||c|c}
& $\kappa=0.7$ & $\kappa=1.0$ \cr
\hline
$\Delta E_{\rm scat.}$      &-0.26264 &-0.26219 \cr
$E_{\rm FD}+E_{\rm CT}$     &-0.01574 &-0.01620\cr
$\Delta E$                  &-0.27839 &-0.27839
\end{tabular}}
\caption{\label{tab:fb}Dependence of components of the VPE on the fake boson parameter $\kappa$ for
$n=4$. Left panel: $D=2+1$, right panel: $D=3+1$.}
\end{table}

Our results for the scattering data contributions $\Delta E_{\rm scat.}$ generally agree with those of
Ref.~\cite{Alonso-Izquierdo:2016bqf}, in particular on the sign and the tendency with increasing 
winding number $n$. Those authors employed a truncated heat kernel expansion with $\zeta$-function 
regularization. This amounts to an $\overline{\rm MS}$ renormalization scheme and we have thus compared 
their results to $\Delta E_{\rm scat.}$. 
We do not agree with the sign in the prediction presented\footnote{The erratum to Ref.~\cite{Baacke:2008sq}
has a sign change compared to the original publication.} in Ref.~\cite{Baacke:2008sq} for $n=1$ and $D=3+1$. 
As in that study we reproduce the significant cancellation between scattering and Feynman diagram 
contributions. Ref. \cite{Baacke:2008sq} constructs a local density of states using the Green's function 
at coincident points, which in turn relies on scattering data along the imaginary momentum axis; this
approach is essentially equivalent to ours, because the integral over space of the Green's function 
yields the Jost function that we compute. Apart from the different sign we also find that our results 
are smaller in magnitude. This may be due to the angular momentum extrapolation, since as we have seen 
the extrapolation from $L\approx35$ they use is likely insufficient \cite{Graham:2019fzo}. In addition, 
the Green's function approach requires an additional integral over the radial coordinate, which may be a 
source of numerical inaccuracies. As a final source of the disagreement, we note that Ref. \cite{Baacke:2008sq} 
imposes the $\overline{\rm MS}$ renormalization scheme at the scale of the un-renormalized gauge boson mass.
In an earlier proof of concept investigation of the $D=3+1$ case we used a simplified on-shell
scheme \cite{Graham:2022rqk}. Comparison reveals that different schemes can easily change the 
sign of the small $n=1$ VPE. However, they do not alter the (almost) linear dependence on the 
winding number.

\section{Conclusions}

We have computed the one-loop quantum corrections to the energy (per unit length) of 
ANO vortices in scalar electrodynamics with spontaneous symmetry breaking, in the BPS case 
where the classical masses of the scalar and gauge fields are equal. These corrections arise from 
the polarization of the spectrum of quantum fluctuations in the classical vortex background.  
This vacuum polarization energy is small because the small coupling approximation applies 
to electrodynamics with $e^2=4\pi/137\approx0.09$, but it becomes
decisive in the case of observables for which the classical result vanishes,
such as the binding energies of vortices with higher winding numbers
in the BPS case.

After clarifying a number of technical and numerical subtleties, we found that the dominant 
contribution to vacuum polarization energies of vortices stems from the non-perturbative
contribution, which cannot be computed from the lowest order Feynman
diagrams; these
diagrams represent an expansion in the background fields rather than the coupling constant. 
Our numerical simulations for vortices with winding number up to four suggest that the quantum
energy weakly binds higher winding number BPS vortices. We have also seen that the vacuum 
polarization energy for the unit winding number vortex is very small, so that at first glance it
appears to be compatible with zero up to numerical errors. The potentially most important source 
for such errors is the small radius behavior in channels that contain zero angular momentum 
components. However, our numerical analysis suggests that any improvement of the data is small 
and likely to push that vacuum polarization energy further away from zero.

To our knowledge these are the first studies of a static soliton vacuum polarization energy that 
compare different topological sectors in a renormalizable model. The vortex model has 
two nontrivial space dimensions. Solitons in one space dimension do either not have topologically 
distinct\footnote{There are different topological sectors in these models but the corresponding 
solitons solutions are constructed from those with the lowest nonzero winding number. For example, the 
$n=2$ sine-Gordon solution is the superposition of two (infinitely) widely separated $n=1$ solutions.}
static solitons, such as the kink and sine-Gordon soliton \cite{Ra82}, or are destabilized by the 
quantum corrections, such as the $\phi^6$ model soliton \cite{Weigel:2019rhr}. The Skyrme model
\cite{Skyrme:1961vq} in three space dimensions indeed has static solitons  with different winding 
numbers, but unfortunately that model is not renormalizable.

For $D=2+1$ and $D=3+1$ the VPE (approximately) decreases linearly with the winding number $n$ with
some offset at $n=0$. For the binding energies $\Delta E-n\Delta E\big|_{n=1}$ we get 
$-0.297(n-1)-0.035(n-1)^2$ and $-0.070(n-1)$, respectively. Since in the BPS case the classical 
binding energy is strictly zero, this implies that it is energetically favorable for vortices
to coalesce rather than to appear in isolation. This observation is characteristic of a type~I superconductor.
We also observe that $\Delta E\big|_{n=4}-2\Delta E\big|_{n=2}<0$ and 
$\Delta E\big|_{n=3}-\Delta E\big|_{n=2}-\Delta E\big|_{n=1}<0$, making the existence of
substructures unlikely.

Away from the BPS case, the classical binding energy can quickly overwhelm the quantum correction 
since the model is weakly coupled. Nevertheless, the computation of vacuum polarization energies
for unequal masses would be desirable to complete this picture.  Technically this calculation is more 
involved because it corresponds to a full $4\times4$ scattering problem, rather than a doubled $2\times2$ 
problem. Another interesting question is whether the techniques to avoid divergences in the Fourier 
transform of the vortices that was developed in Ref.~\cite{Graham:2019fzo} and employed here will 
also be successful when coupling fermions and thus avoid the necessity
of a {\it return flux} 
\cite{Graham:2004jb}. If this is the case, supersymmetric extensions \cite{Edelstein:1993bb} can 
be investigated as well. We also conjecture that analogous calculations are possible in the case of a 
't Hooft-Polyakov monopole \cite{tHooft:1974kcl,Polyakov:1974ek}.
\bigskip

\acknowledgments
N.\@ G.\@ is supported in part by the National Science Foundation (NSF)
through grants PHY-1820700 and PHY-2205708.
H.\@ W.\@ is supported in part by the National Research Foundation of
South Africa (NRF) by grant~109497.

\appendix

\section{Higher order effects from the wave-function renormalization of the gauge field}
\label{appA}

In this Appendix we briefly discuss the higher order effects of
$\mu\ne1$. These arise via the wave-function counterterm for the gauge field in
Eq.~(\ref{eq:Lct}). The changes in the energies are
\begin{align}
E_\mu&=-\frac{M}{96}\left[\frac{12}{4-\mu^2}+\frac{6}{\mu}{\rm atanh}\left(\frac{\mu}{2}\right)
-4-3\ln(3)\right] \int_0^\infty \rho d\rho\left(n\frac{g^\prime}{\rho}\right)^2\,, \cr
E_\mu&=\frac{M^2}{288\pi}\left[\frac{24}{\mu^2}\left(1-\mu^2\right)+5\sqrt{3}\pi
-\frac{6}{\mu^3}\frac{\mu^4-2\mu^2+16}{\sqrt{4-\mu^2}}{\rm asin}\left(\frac{\mu}{2}\right)\right]
\int_0^\infty \rho d\rho\left(n\frac{g^\prime}{\rho}\right)^2
\label{eq:Emu1}\end{align}
for $D=2+1$ and $D=3+1$, respectively. These changes are the differences of the gauge field counterterm 
contributions evaluated at $\mu$ and $\mu=1$. Interestingly, we do not need to explicitly solve 
Eqs.~(\ref{eq:consit2}) and~(\ref{eq:consit3}) to determine $\mu$ for given values 
of $e$ and $M$, as long as we measure the energies in units of $M$ and~$M^2$. However, before we
compute $E_\mu$, we must identify the range of solutions to these
equations. We therefore write them as 
\begin{align}
0&=1-\mu^2+\zeta\left[\frac{15}{8}\ln(3)-\frac{13}{6}
+\frac{\mu^2}{2\left(4-\mu^2\right)}-\frac{\mu}{4}{\rm atanh\left(\frac{\mu}{2}\right)}\right]\cr
0&=1-\mu^2+\zeta\left[\frac{9}{2}-\frac{5\pi}{\sqrt{3}}-\frac{\mu^2}{6}
+\frac{16-2\mu^2}{\mu\sqrt{4-\mu^2}}{\rm asin}\left(\frac{\mu}{2}\right)\right]\,,
\label{eq:Emu2}\end{align}
with the dimensionless parameters $\zeta=\frac{e^2}{2\pi M}$ and $\zeta=\frac{e^2}{8\pi^2}$ for $D=2+1$ 
and $D=3+1$, respectively. The singularity at $\mu=2$ emerges from virtual Higgs 
particles going on-shell and allowing the gauge particle to decay into
two Higgses. Even if this singularity
produces a zero crossing Eq.~(\ref{eq:Emu2}), we consider it unphysical because it does not smoothly emerge from the tree-level
result. Such a solution would approach $\mu=2$ as $\zeta\,\to\,0$.

\begin{figure}
\centerline{\epsfig{file=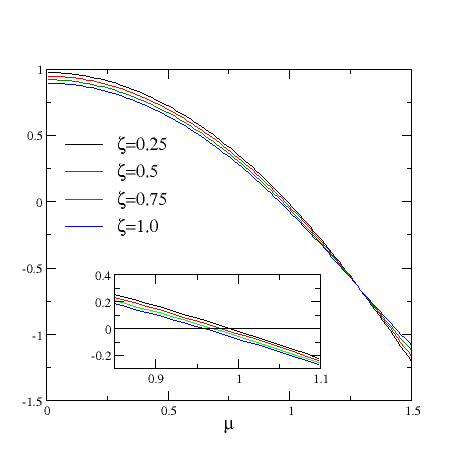,width=7cm,height=4cm}~~~
\epsfig{file=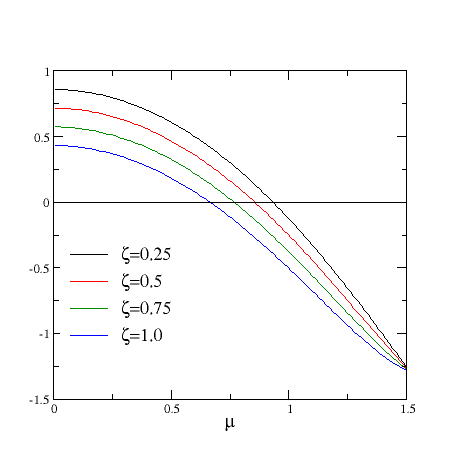,width=7cm,height=4cm}}
\caption{\label{fig:appendix}Right-hand sides for the implicit Eqs.~(\ref{eq:Emu2}).
Left panel: $D=2+1$, right panel: $D=3+1$.}
\end{figure}

We display the right-hand sides of Eq.~(\ref{eq:Emu2}) in Figure \ref{fig:appendix}.
For $D=2+1$, only a narrow window below $\mu=1$ is accessible even for un-realistically large values
of $\zeta$. On the other hand, for $D=3+1$ it seems acceptable to have $\mu\in[0.8,1]$.
In Tables \ref{tab:appendix1} and \ref{tab:appendix2} we present the numerical results for 
$E_\mu$ in the appropriate ranges of $\mu$.
\begin{table}
\begin{tabular}{c|cccc}
$\mu$ & $n=1$ & $n=2$ & $n=3$ & $n=4$ \cr
\hline
0.92  & 0.0017 & 0.0028 & 0.0048 & 0.0069\cr
0.94  & 0.0008 & 0.0021 & 0.0036 & 0.0053\cr
0.96  & 0.0006 & 0.0015 & 0.0025 & 0.0036\cr
0.98  & 0.0003 & 0.0007 & 0.0013 & 0.0018\cr
\end{tabular}
\caption{\label{tab:appendix1}
Gauge mass variation of the VPE, $E_\mu$, for $D=2+1$ in units of $M$ according to Eq.~(\ref{eq:Emu1}).}
\end{table}
\begin{table}
\begin{tabular}{c|cccc}
$\mu$ & $n=1$ & $n=2$ & $n=3$ & $n=4$ \cr
\hline
0.80 & 0.0004 & 0.0011 & 0.0018 & 0.0027\cr
0.85 & 0.0003 & 0.0009 & 0.0014 & 0.0021\cr
0.90 & 0.0002 & 0.0006 & 0.0010 & 0.0015\cr
0.95 & 0.0001 & 0.0003 & 0.0005 & 0.0008\cr
\end{tabular}
\caption{\label{tab:appendix2}
Gauge mass variation of the VPE, $E_\mu$, for $D=3+1$ in units of $M^2$ according to Eq.~(\ref{eq:Emu1}).}
\end{table}
Compared to the data in Tables \ref{tab:D2} and \ref{tab:D3}, these
contributions are tiny and do not effect our conclusions on the VPE.

\bibliographystyle{apsrev}

\end{document}